\documentclass[preprint]{vgtc}               
\usepackage{graphicx}
\usepackage{amsmath,amssymb,mathtools}
\usepackage{booktabs}
\usepackage{enumitem}
\usepackage{microtype}
\usepackage{listings}
\usepackage{xcolor}
\graphicspath{{figures/}{./}}

\lstdefinestyle{py}{
  language=Python,
  basicstyle=\ttfamily\small,
  columns=fullflexible,
  breaklines=true,
  frame=single,
  showstringspaces=false,
  tabsize=2
}

\vgtcpapertype{theory/model}
\vgtccategory{Research}

\title{\textbf{Beyond the Post Hoc User Study}: Modeling Visual Decision-Making with Active Inference}

\author{%
  Harrison J.~Goldwyn,
  Graham Johnson,
  Christopher Ibarra,
  Lace Padilla,
  and Kenny Gruchalla
}

\authorfooter{%
  \item Harrison J.~Goldwyn (harrison.goldwyn@nlr.gov), Graham Johnson, Christopher Ibarra, and Kenny Gruchalla are with the National Laboratory of the Rockies (NLR).
  \item Lace Padilla is with Northeastern University
}

\abstract{The evaluation of visual encodings is grounded in empirical user studies, including controlled comparisons of alternative designs. 
These studies are essential for measuring whether a visualization supports accurate human judgment, and many can provide evidence about underlying perceptual and cognitive mechanisms. 
However, empirical evidence alone does not enable causal prediction of interpretation errors.
In this sense, evaluations are often post hoc: they assess visualization efficacy after a design has been specified, rather than predicting how human attention, uncertainty, memory, and bias may lead a viewer to accurate or erroneous interpretation. 
This mechanistic gap 
restricts the field’s ability to accumulate predictive design knowledge. 
Even with evidence-supported cognitive frameworks for visual decision-making, we 
lack general 
means to simulate user behavior and predict errors \emph{in silico}. 
To bridge this gap, 
we demonstrate
a translation of a cognitive theory of visualization interpretation into executable simulation
using Active Inference: a probabilistic process theory of living systems often applied to human perception, learning, and action-taking. 
Active Inference agents iteratively minimize the probability of surprising observations by updating internal belief states and choosing informative actions. 
Simulating human interpretation of data visualization in this context, we frame chart reading as a dynamic visual search, minimizing both uncertainty 
and cognitive effort 
to reach task completion. 
As a foundational proof-of-concept, we engineer Active Inference agents that perform a bar-chart average-estimation task under a dual-process theory of decision-making. 
Crucially, our architecture aims to replicate human perceptual vulnerabilities by presenting a Fast, heuristic (Type 1) agent prone to tick salience bias and a Slow, analytic (Type 2) agent more prone to working-memory decay. 
Both agents yield inspectable cognitive traces, including the evolution of belief uncertainty and the chosen visual fixation sequences. 
By distilling these hypothesized failure mechanisms into interpretable parameters, 
we present this architecture  
as a framework for formalizing hypotheses about mechanisms of visualization interpretation. 
These models highlight a new role for empirical studies in visualization design: to parameterize, test, refine, or falsify decision-making simulations. 
Fitting these models to user data could support a shift from 
primarily post hoc testing toward predictive \emph{in silico} validation, allowing researchers to anticipate efficacy 
earlier in the design process.
}

\keywords{Perception and Cognition, Visualization Design and Evaluation, Cognitive Modeling, Active Inference, Decision Making}

\CCScatlist{%
  \CCScat{Human-centered computing}{Visualization}{Visualization theory, concepts and paradigms}{}
  \CCScat{Computing methodologies}{Artificial intelligence}{Cognitive robotics}{}
}

\teaser{
  \centering
  \includegraphics[width=\linewidth]{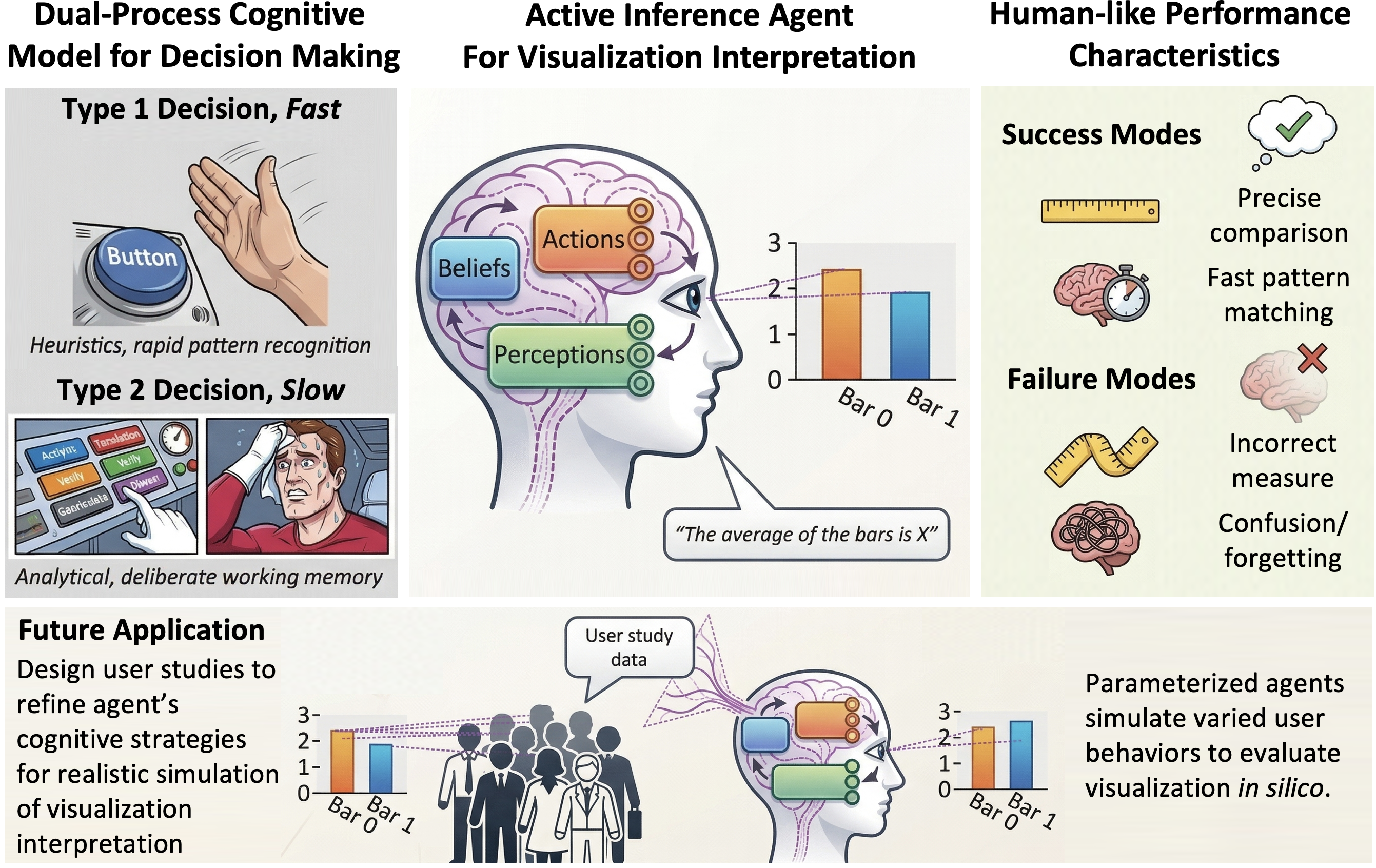}
  \caption{%
  	Diagram covering the integration of a Dual-Process theory for decision making with Active Inference for simulating human-like interpretation of data visualization. The top-left panel illustrates the Type 1, \emph{Fast}, process and Type 2, \emph{Slow}, process. The top-middle panel illustrates an Active Inference agent viewing a bar chart and determining the average bar value. The top-right panel describes the interpretable human-like success and failure modes an Active Inference agent can instantiate. The bottom panel describes our ultimate call-to-action: parameterizing Active Inference agents with user studies for \emph{in silico} evaluation of visualization efficacy.}
  \label{fig:teaser}
  \vspace{-5pt}

}

\begin{document}
\setlength{\abovedisplayskip}{1pt}       
\setlength{\belowdisplayskip}{1pt}       

\firstsection{Introduction}
\maketitle

\subsection{The need for mechanistic modeling}
The field of visualization is built upon a rich, hard-won foundation of empirical research. 
Decades of perceptual psychology and rigorously controlled user studies have yielded robust guidelines for visual design, mapping out hierarchies of visual encodings and perceptual accuracies~\cite{franconeri2021science}. 
However, 
empirical evaluation alone does not enable prediction of design efficacy. 
Controlled comparisons can reveal that one design produces higher error rates than another, but they do not necessarily provide a computational account of how those errors arise within the viewer's cognitive process.
By ``post hoc'' evaluation, we refer to studies that assess performance after a visualization design has already been specified or implemented.
Such studies remain indispensable; controlled perceptual experiments, eye tracking, response-time studies, think-aloud protocols, and process tracing
can all provide evidence about the mechanisms underlying visualization interpretation.
But such evidence without a process model can leave designers in an expensive trial-and-error loop without a predictive account of the causal pathways that generate error.

To understand the \emph{how} of visualization failures, the visualization community has increasingly turned to cognitive science~\cite{szafir2023visualization}. Frameworks such as Padilla et al.'s account of decision making with visualizations have provided evidence for specific conceptual pathways of visual data extraction \cite{Padilla2018DecisionMaking}. In particular, they highlight the distinction between Type~1 processing (fast, heuristic, perceptual judgments) and Type~2 processing (slow, effortful, analytic graph reading)~\cite{kahneman2002representativeness, stanovich1999rational, Evans2013DualProcess}. 
While such frameworks provide invaluable 
conceptual accounts, they are not yet executable process models.
A visualization designer cannot currently input a proposed chart into these frameworks and simulate whether a user is likely to succeed with a fast heuristic, fall into a predictable bias, or be forced into a slower and more demanding analysis.

Visualization interpretation is not a passive reception of pixels, but an active process of perceptual sampling, belief updating, and decision commitment \cite{pinker2014theory, hegarty2011cognitive, shah2005comprehension, Padilla2018DecisionMaking, brumar2026typologydecisionmakingtasksvisualization}. Users must continually balance the drive to gather information and minimize uncertainty against the need to avoid cognitive overload, shifting between rapid heuristic judgments and slower analytic computations depending on visual clarity. In this paper, we argue that this cognitive balancing act can be mathematically formalized within Active Inference \cite{friston2006free, Friston2017ActiveInference}. By representing chart reading as action and inference in a partially observable environment, Active Inference provides a principled modeling framework for latent task states, observations, actions, preferences, and memory dynamics. 
This approach allows us to close the gap between qualitative cognitive theory and computable mechanism, 
moving from conceptual mechanism descriptions toward executable simulations that generate explicit, testable predictions about why particular errors arise.

We do not suggest that synthetic cognitive agents will replace human subjects in visualization evaluation. Rather, we view them as a mechanistic scaffold for a more generative science of visualization design: a way to formalize hypotheses about attention, memory, and decision-making, and to test how these mechanisms interact with specific visual encodings. The immediate limitation of this prototype is the need for empirical grounding. This is therefore also a call to action: by leveraging eye tracking, reaction times, error distributions, and related behavioral data to parameterize and validate such models, the field can
use empirical data not only to evaluate completed designs, but also to parameterize, test, and refine computational simulations of where and why a visualization is likely to succeed or fail.

\subsection{From qualitative cognitive models to simulation}

\begin{figure}[t]
  \centering
  \includegraphics[width=\linewidth]{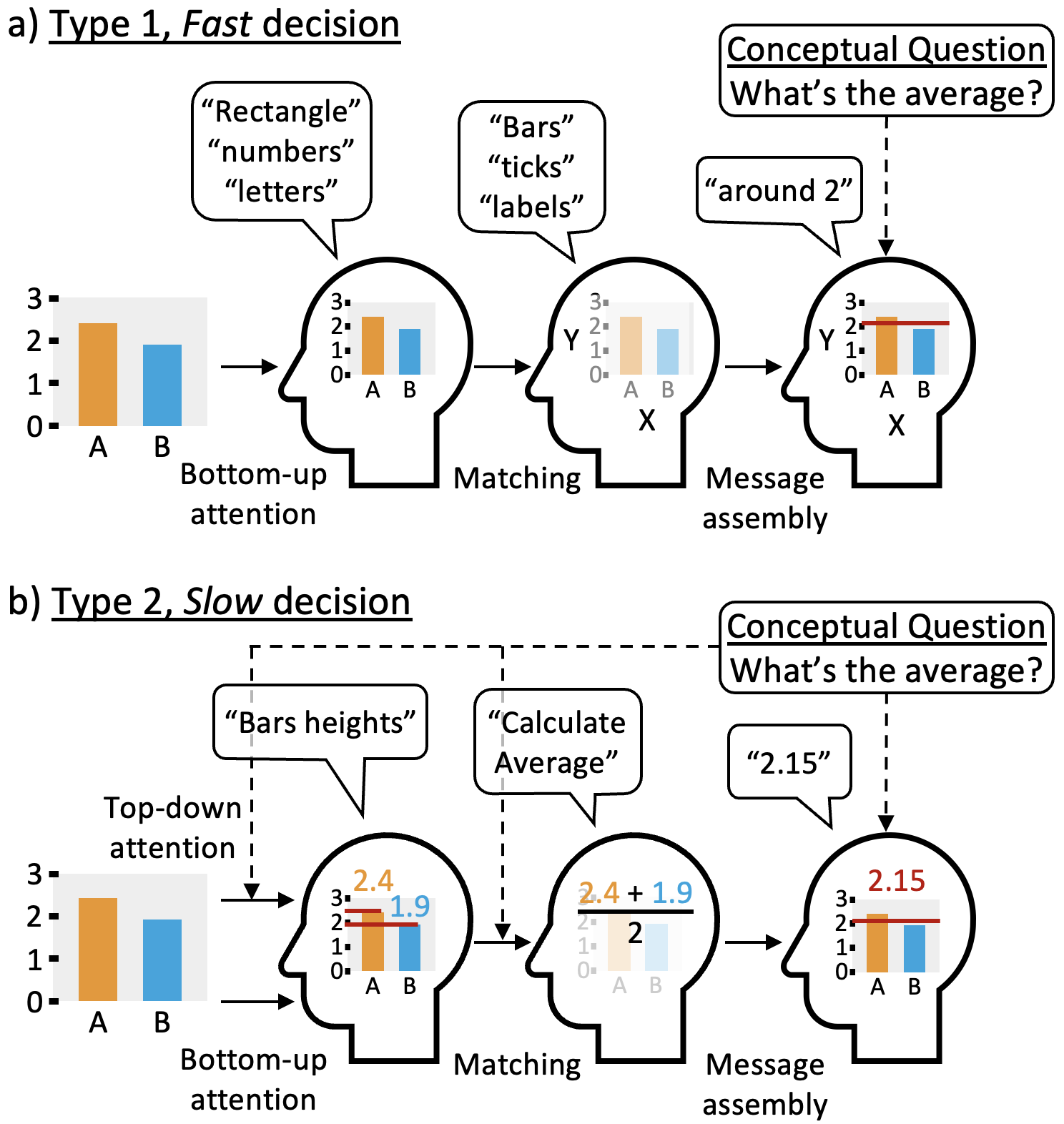}
  \caption{An illustration of the Dual-process decision making proposed by  Padilla~et~al.~\cite{Padilla2018DecisionMaking}. Type~1 (top) uses a fast heuristic decision with minimal working-memory involvement, while Type~2 (bottom) performs a more analytic computation that uses working memory.}
  \label{fig:padilla}
\end{figure}

A particularly influential qualitative account of visualization-aided decision making is the cognitive framework of Padilla~et~al.~\cite{Padilla2018DecisionMaking}. Synthesizing evidence across domains, Padilla~et~al.\ argue that decisions with visualizations can be understood through a dual-process lens. \emph{Type~1} decisions are fast and heuristic, often relying on coarse perceptual impressions with minimal working-memory demand. \emph{Type~2} decisions are slower and more analytical, recruiting significant working memory to perform intermediate computations and to support accuracy under uncertainty (cf.\ broader dual-process theory \cite{kahneman2002representativeness, stanovich1999rational,Evans2013DualProcess}). Figure~\ref{fig:padilla} shows their canonical example: deciding whether the average of two bars is closer to $2$ or $2.2$. Even in this simple setting, decades of work in graphical perception and graph comprehension indicate that performance reflects structured operations such as feature extraction, reference-frame alignment, and memory for intermediate estimates \cite{Cleveland1984GraphicalPerception,pinker2014theory}.

Our central hypothesis is that Active Inference can render key dimensions of this qualitative account mathematically explicit and executable. 
In particular, Padilla~et~al.'s framework suggests that visualization judgments vary along dimensions such as representational compression, working-memory demand, cognitive effort, and sequential control.
If 
these dimensions capture
capture
relevant cognitive structure of the task, then an Active Inference instantiation should reproduce not only correct responses, but also qualitatively human-like \emph{processes} and \emph{failure modes}. 
In this work, we therefore treat Padilla~et~al.'s framework as a target description of 
two idealized task strategies: compressed heuristic estimation (Type 1, Fast) and sequential analytic estimation (Type 2, Slow).
Although the separability of these processes in human cognition remains debated \cite{keren2009two,melnikoff2018mythical}, this idealization lets us examine the consequences of making each strategy computationally precise: the beliefs it forms, the actions it selects, and the failure modes it predicts under parameterized perceptual bias or memory decay.

As a foundational proof-of-concept, we focus on the canonical visualization task illustrated in Figure \ref{fig:padilla}: estimating the average of two bars in a bar chart. 
Our core claim is not that this model predicts visualization failures in general nor do we propose a full computational implementation of dual-process cognition.
Instead, our aim is to show that Active Inference can instantiate hypothesized cognitive mechanisms for a constrained chart-reading task and produce interpretable, mechanism-specific failure patterns.
To test this, we construct Type~1 (\emph{Fast}) and Type~2 (\emph{Slow}) agents from idealized Fast and Slow strategies for average estimation: compressed evaluation of the visual bar-pair midpoint and sequential bar-wise height estimation with intermediate memory. 
To compare models within a common computational framework, we use identical parameters wherever correspondence is appropriate and allow parameters to differ only in the architectural and temporal features required by their respective cognitive modes. 
This matched parameterization enables comparison in a controlled computational experiment instead of arbitrary contrast of unrelated models.

\subsection{Summary of contributions}

The primary contribution of this paper is a worked example of a mechanistic modeling methodology for visualization interpretation. Rather than proposing a new theory of visual cognition or replacing empirical evaluation, we show how an existing cognitive theory can be translated into inspectable Active Inference agents that generate testable process-level predictions. Concretely, we demonstrate this through:
\begin{itemize}
    
    
    \item \textbf{Dual-Process Example Architecture:} We implement Padilla et al.'s Type~1 (\emph{Fast}) and Type~2 (\emph{Slow}) cognitive framework as discrete-time Active Inference agents solving Partially Observable Markov Decision Processes (POMDPs) on a bar-chart average-estimation task. We demonstrate how a hypothesized cognitive strategy can be used to specify latent task variables, perceptual observations, action policies, cognitive biases and limitations.

    \item \textbf{Inspectable Failure Predictions:} We show how models encoded with human-like perceptual vulnerabilities, such as tick-salience bias and working-memory decay, generate distinct error patterns, belief trajectories, and fixation sequences. Future empirical studies can use these outputs to falsify or support the hypothesized cognitive strategy underlying model construction.


\end{itemize}


\noindent Together, these elements establish a proof of concept for executable cognitive modeling in visualization. Generalizing this approach to arbitrary encodings and broader interactive decision-making tasks will require richer perceptual models, larger policy spaces, parameter fitting, and validation against human behavioral data.

\paragraph{Scope.}
We focus on the \emph{cognitive} and \emph{decision} layers of visualization use: attention allocation, memory, inference, and reporting. Perceptual front-ends that convert raw pixels into symbolic chart primitives (e.g., computer vision segmentation and labeling pipelines) are beyond the scope of this paper and are left for future work (Section~\ref{sec:future}).

\section{Background}

\subsection{Visualization cognition and dual-process accounts}

To contextualize Padilla et al.’s framework, we briefly review computational modeling of visualization interpretation to highlight what existing approaches capture well and where Active Inference may provide a useful modeling perspective.

Graphical perception studies quantify which encodings support accurate judgments, such as position versus length~\cite{Cleveland1984GraphicalPerception}. Cognitive accounts of graph comprehension emphasize that viewers build structured internal representations from visual marks and perform task-driven inferences over those representations~\cite{pinker2014theory}. Padilla~et~al.~\cite{Padilla2018DecisionMaking} connect these traditions to dual-process decision theory~\cite{Kahneman2011ThinkingFastSlow,Evans2013DualProcess}, arguing that visualization judgments often default to fast heuristic processes and recruit slower analytic processes when task demands or incentives justify additional cognitive effort. Together, these perspectives clarify \emph{what} kinds of perceptual and cognitive operations matter for visualization use, but they do not by themselves provide a unified computational account of how visual sampling, memory, task goals, and decision commitment interact over time. Moreover, recent work has highlighted how the types of decision making tasks can be taken into account when designing decision-support tools~\cite{brumar2026typologydecisionmakingtasksvisualization}.

Several computational traditions address parts of this problem. Top-down cognitive architectures, such as ACT-R \cite{Anderson2004AnIT}, have been highly successful in modeling working memory and rule-based problem solving, but they often abstract away the continuous, noisy nature of visual perception. Conversely, bottom-up saliency models \cite{ittiSaliency, bylinskii2017differentevaluationmetricstell} excel at predicting where a user will look based on pixel-level features, but they struggle to account for how high-level task goals override visual salience. Information Foraging Theory \cite{pirolliForaging} models how users seek information to maximize value, but it is typically applied at the macro-level of navigating websites or document spaces rather than the micro-level of intra-chart visual saccades.

Recent work has emerged around the concept of the ``virtual viewer'', modeling visualization interpretation with primary focus on prediction of user eye movement within task-driven chart reading.
Systems such as A Scanner Deeply and Perceptual Pat predict gaze heatmaps or provide automated perceptual critiques of visualization designs, supporting the broader idea that computational models can complement empirical evaluation during design iteration \cite{borkin2022scannerDeeply,shin2023perceptualPat}. More recent models move from aggregate attention maps toward task-conditioned visual behavior. Wang et al.\ propose a unified model of saliency and scanpaths for information visualizations, showing that temporal fixation sequences provide information not captured by static saliency alone \cite{wang2021scanpathVisualization}. SalChartQA further demonstrates that visual attention is strongly conditioned by the viewer's information need, introducing a large question-driven saliency dataset and a transformer model for predicting saliency from visualization--question pairs \cite{wang2024salchartqa}. Most directly, Chartist simulates task-driven scanpaths for chart-reading tasks such as value retrieval, filtering, and finding extrema using a hierarchical controller that combines high-level task interpretation with reinforcement-learned eye-movement policies \cite{shi2025chartist}. These approaches establish that task context strongly shapes where viewers look during chart interpretation. 
Our work is a mechanistic approach to the ``virtual viewer'': rather than predicting gaze, saliency, or scanpaths directly, we use Active Inference to specify an explicit generative model of the latent cognitive process by which an agent updates beliefs based on visual information, resolves uncertainty, and commits to a response.

More recently, Bayesian cognitive models have made significant inroads into graphical perception, formalizing how users combine prior knowledge with noisy visual evidence \cite{2019-bayesian-cognition-vis}. These models are important because they make latent perceptual uncertainty explicit. However, many such models treat perception as a single inference step over a fixed stimulus. They do not directly model chart reading as a sequential control problem in which the viewer actively decides where to look to resolve uncertainty, updating memory over time, and eventually committing to a task response.

The critical gap in the visualization literature is the lack of a unified computable framework that integrates top-down task goals, bottom-up sensory noise, working memory, and sequential information-seeking into a single inspectable mechanism and objective function. We require an architecture that express both rapid heuristic approximations and effortful analytic sequences, while exposing the latent beliefs and uncertainty dynamics that lead to success or failure. Active Inference provides this framework by treating visualization interpretation as sequential perception and action under uncertainty.

\subsection{The Active Inference Formalism}

Active Inference is a probabilistic modeling framework for the dynamics of living systems. It was developed by computational neuroscientists wishing to model human cognition and behavior \cite{Friston2017ActiveInference}. Active Inference agents iteratively minimize the probability of surprising observations by adjusting internal belief states and choosing informative actions. This process treats perception and action as two sides of the same coin: the drive to reduce unexpected and undesired outcomes. Unlike traditional visualization models that treat the viewer as a passive processor of visual input, Active Inference models the viewer as an active agent.

The framework is grounded in the idea that the role of the brain is to predict environmental state from sensory observation. These predictions are made by maintaining, updating, and testing a generative model. This model is ``generative'' in the sense that it predicts the evolution of environmental state and how these true world-states generate observations. Better understood as a set of axioms for model construction rather than a single process model, Active Inference is extremely flexible and therefore notoriously difficult to understand and implement \cite{mann2022free, smith2022step}. 
This has led to the literature being flush with tutorials and alternative theoretical perspectives. Here, we introduce the theory only to convey the required elements of model construction. 

\begin{figure*}[!t]
    \centering
    \includegraphics[width=0.8\textwidth]{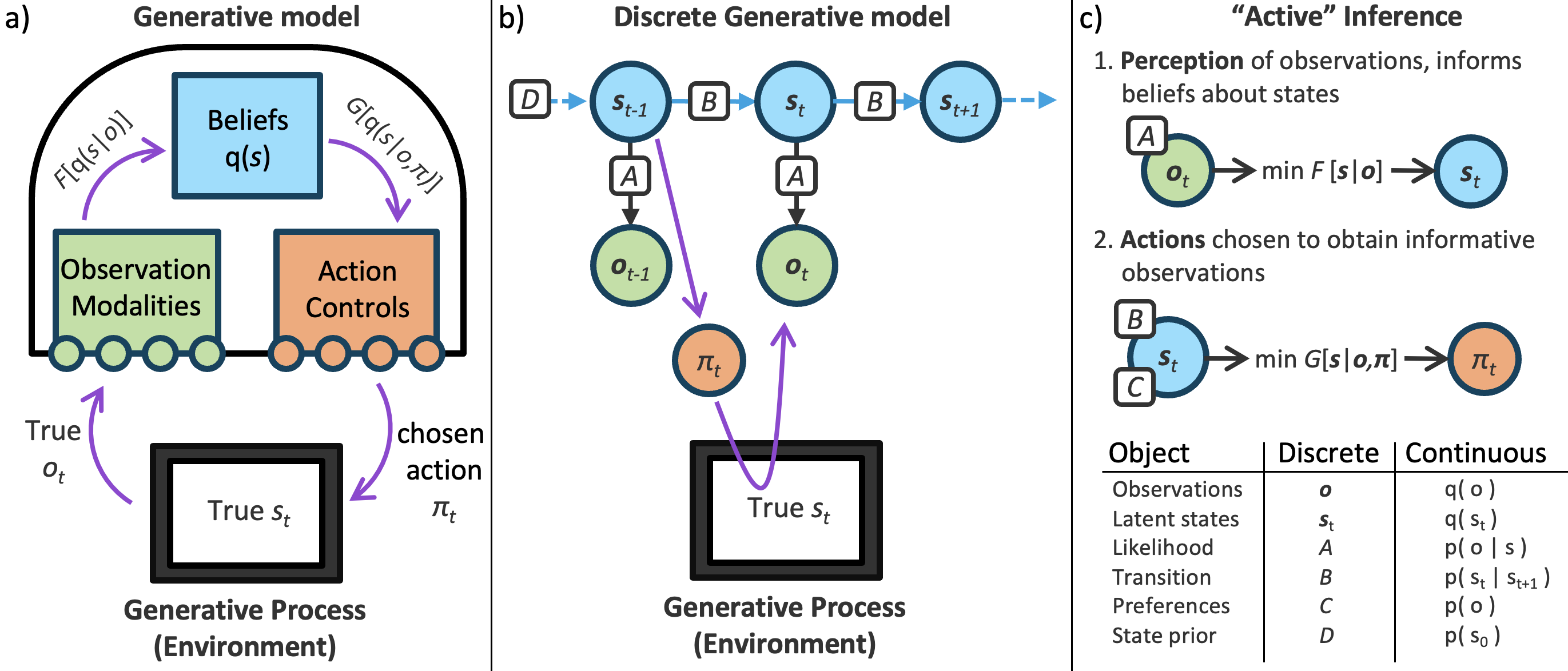}
    \caption{The Active Inference Formalism. Panel (a) illustrates the interaction between the agent (generative model) and the environment (generative process). The generative model is updated by selecting actions which are optimally informative under the Expected Free Energy $G$. Chosen actions $\pi$ (or sequences of actions) are the agents means of interacting with the environment. In response, the agent receives observations $o$, which are used to update its beliefs about environmental states $q(s)$ by minimizing Free Energy $F$.
    Panel (b) illustrates a discrete time implementation of the generative model. The agent relates perceived observations to its beliefs about states through the likelihood matrix $A$ and predicts the dynamics of environmental states through the transition matrix $B$.
    Panel (c) describes the perception and action-selection processes fundamental to Active Inference and highlights discrete-time notation and mathematical dependencies. 
    }
    \label{fig:actinf}
\end{figure*}

The first of two axioms is called the \emph{Bayesian Brain Hypothesis}. 
To introduce this axiom, we first consider that living organisms are separate from their environment in the sense that they only obtain indirect information about the world through their sensory organs. 
The Bayesian Brain Hypothesis suggests that the role of the brain is to construct a probabilistic model of reality based on sensory perception. 
By nature of being alive, organisms are not simply passive sensors but \emph{agents} who can utilize their internal model to take action based on their predictions about the environment to provoke informative or rewarding sensations.  
Figure \ref{fig:actinf} illustrates this perspective on organism/environment separation.
The agent's generative model must be probabilistic to meet the needs of living systems to predict probable futures and likely outcomes of their actions. 
Because agents must infer environmental states from their sensory input, we describe the brain's generative model as probabilistic beliefs over future observations $o$ and states $s$.

The second axiom is the \emph{Free Energy Principle} for the Brain \cite{friston2006free}. The Free Energy principle proposes the mathematical rules for updating the generative model based on sensory input and guides the agent in action selection. The fundamental idea is that an organism's ultimate goal is to maintain some homeostasis. To do so, it must minimize unexpected and unwanted sensations (i.e., observations). Mathematically, we can frame this as the minimization of Bayesian surprise:
\begin{equation}
    -\ln p(o_t) = -\ln \sum_{s_t} p(o_t,s_t)
\end{equation}
Because the sum over all possible environmental states required to compute the surprise from the generative model is generally intractable, we can equivalently minimize an expression called the \emph{Variational Free Energy}, \textbf{$F$}, named for its similar form to thermodynamic free energy \cite{friston2006free, smith2022step}. 
The definition of Free Energy introduces an approximate posterior \(q_t(s_t) \approx p(s_t \mid o_t)\), which opens up several mathematical conveniences discussed extensively in the literature \cite{parr2022active}. 
Thus, minimizing variational free energy both minimizes an upper bound on surprise and brings the approximate posterior \(q_t(s_t)\) closer to the exact posterior \(p(s_t\mid o_t)\). A representation of Variational Free Energy that is useful for interpretation is: 
\begin{equation}
F[q_t]
=
\underbrace{-\mathbb E_{q_t(s_t)}\!\big[\ln p(o_t\mid s_t)\big]}_\text{inaccuracy}
+
\underbrace{D_{\mathrm{KL}}\!\big(q_t(s_t)\,\|\,p(s_t)\big)}_\text{complexity}
\label{eq:free_energy_accuracy_complexity}
\end{equation}
In minimizing variational free energy, the agent minimizes \emph{inaccuracy} and \emph{complexity}. The \emph{inaccuracy} term penalizes beliefs $s_t$ that poorly explain the current observation $o_t$, while the \emph{complexity} term penalizes posterior beliefs $q_t(s_t)$ that deviate from prior beliefs $p(s_t)$.

The last important implication of the Free Energy Principle is how the agent selects actions by extending the same logic into the future.
Because future observations have not yet occurred, an agent cannot directly minimize a realized surprise.
Instead, it evaluates candidate policies as series of potential actions under \emph{Expected Free Energy}, \textbf{$G$}, which scores the future states and outcomes that each policy is expected to induce. Following conventional notation, policies are written as \(\pi = (u_t,u_{t+1},\dots,u_{t+H-1})\) over a planning horizon \(H\).
For a future time \(\tau \ge t\), let \(q(s_\tau \mid \pi)\) denote the predictive belief over hidden states under policy \(\pi\), and let \(q(o_\tau,s_\tau \mid \pi)\)
denote the corresponding predictive joint distribution over future states and outcomes.
We write \(p^\ast(o_\tau)\) for the agent's prior preference distribution over future outcomes.
Using this notation, an instructive decomposition of expected free energy is
\begin{multline}
G(\pi)
=
\sum_{\tau=t}^{t+H-1}
\Big[
\underbrace{
\mathbb E_{q(o_\tau \mid \pi)}
\big[
-\ln p^\ast(o_\tau)
\big]
}_{\text{pragmatic cost / preference violation}}
\\
-
\underbrace{
\mathbb E_{q(o_\tau,s_\tau \mid \pi)}
\!\left[
\ln q(s_\tau \mid o_\tau,\pi) - \ln q(s_\tau \mid \pi)
\right]
}_{\text{epistemic value / expected information gain}}
\Big]
\label{eq:efe_general}
\end{multline}
The first term encourages outcomes consistent with prior preferences, while the second favors informative observations that are expected to reduce uncertainty about task-relevant hidden states.
Policies are then selected by sampling from (or maximizing)  
$q(\pi) \propto 
\exp\!\big(-\gamma\,G(\pi)\big)$.


This formulation is the crux of our cognitive model. It elegantly balances two competing cognitive drives. The first term, \emph{pragmatic cost}, penalizes policies that lead to outcomes violating the agent's preferences (e.g., prioritizing correct answers, or penalizing cognitive fatigue and excessive time steps). The second term, negative \emph{epistemic value}, quantifies the expected information gain. It drives the agent to take actions, such as saccading to an axis label, that will maximally reduce its uncertainty about the hidden state. 

For the discrete models used in this paper, these ideas are implemented as a Partially Observable Markov Decision Process (POMDP). This gives us a rigorous ledger for the chart-reading process: hidden states encode latent task variables, outcomes encode perceptual evidence and feedback, actions encode shifts of attention and report decisions, and prior preferences encode the agent's goals. Expressed in this form, Active Inference becomes a transparent specification of how uncertainty, memory, and action interact during visualization use.

To enable simulation, we transition from continuous distributions to discrete categorical distributions. This procedure highlights exactly what must be specified to construct an Active Inference agent. 
Figure \ref{fig:actinf} panel B illustrates the discrete generative model in graphical form. We start by specifying independent hidden states into $N_s$ separate \emph{factors} and $N_o$ sensory channels as observation \emph{modalities}. 
Posterior beliefs about hidden states are represented by categorical belief vectors
$\mathbf{s}_t^{(n)}$, with components corresponding to each of $j$ possible values for hidden-state factor $n \in N_o$, \begin{equation}
    \mathbf{s}_t^{(n)}(j) = q(s_t^{(n)} = j).
\end{equation}
Likewise, beliefs about observations in modality $m \in N_m$
are represented by $\mathbf{o}_t^{(m)}$, whose $i$th component is
\begin{equation}
\mathbf{o}_t^{(m)}(i) = q\!\left(o_t^{(m)} = i\right).    
\end{equation}

In canonical discrete-state Active Inference, a generative model specifies a prior over initial states, a likelihood mapping from states to outcomes, controlled state transitions, and prior preferences over outcomes. Using the standard POMDP matrix notation, these components may be written as:
\begin{equation} \label{eq:abcd}
\begin{aligned}
A^{(m)}_{ij} &= p(o_t^{(m)} = i \mid s_t = j), \\
B^{(n)}_{ij} &= p(s_{t+1}^{(n)} = i \mid s_t^{(n)} = j), \\
D_i &= p(s_1 = i), \\
C_i &= p(o_\tau = i).
\end{aligned}
\end{equation}
Here, \(A^{(m)}\) is the likelihood matrix for modality \(m\), \(B^{(u)}\) is the transition matrix induced by action \(u\), \(D\) is the prior over initial hidden states, and \(C_\tau\) encodes preferences over future outcomes. Appendix \ref{app:formalism} details how $F$ and $G$ are computed in this discrete formulation. The minimum specification of an Active Inference agent is simply these pieces: the $N_n$ state factors and their $N_o$ observation modalities along with the matrices $A$, the agent's model for how environmental states generate observations, $B$, the agent's model for how environmental states evolve in time, $C$, the agent's observation preferences, and $D$, the agent's prior expectations of environmental states.

\subsection{Active Inference for Visualization}

With the formal ingredients of Active Inference in place, we can now consider its specific application to human interpretation of data visualization. Recent work in Human-Computer Interaction (HCI) has proposed Active Inference as a broad framework for designing and evaluating user interfaces \cite{murray2025active}. Building on this foundation, we focus specifically on \emph{in silico} evaluation: modeling the user via Active Inference to simulate and interpret interaction behaviors.

In this context, Active Inference serves as a computational bridge that unifies perception, action, and decision-making. Rather than passively reading marks off a page, the agent actively infers data values from noisy visual evidence (perception). Eye movements become epistemic foraging, shifts of attention deliberately chosen to reduce uncertainty about task-relevant quantities (action). Finally, the agent commits to a response only when the expected value of further information no longer outweighs the pragmatic cost of continued sampling (decision). This perspective is appealing for visualization because it integrates visual search, working memory, and response selection into a single generative framework. Within this framework, modelers have significant flexibility in choosing the level of abstraction for latent states and observations. Neurological literature suggests the brain relies on hierarchical generative models, where higher levels encode slower, abstract structures and lower levels govern fast, local dynamics \cite{parr2017working,parr2018anatomy,friston2018deep,palacios2020markov}. This biological hierarchy naturally parallels the dual-process cognitive framework of Padilla \emph{et al.} Consequently, we abstract away the low-level perceptual mechanics (e.g., raw pixel processing) and focus our models entirely on the higher-level cognitive processes. Our agents assume the visual stimulus is already recognized as a bar chart with meaningful components (bars, ticks, \emph{etc.}). We make this deliberate choice based on the hypothesis that the critical errors leading to visualization misinterpretation occur precisely at this layer of relational reasoning and working memory, rather than in basic visual feature extraction. The integration of low-level perception is future work.
 
\section{Methods}
\label{sec:methods}

\subsection{Task: reporting the average}

We study a simple yet cognitively structured visualization task: given a bar chart with two bars and a vertical axis labeled with integer ticks, the observer must report the bars' average value to one decimal place. Although the display is minimal, the task is not trivial. To answer correctly, an observer must map spatial height to axis value, acquire information about the relevant bar heights or their average, maintain intermediate information across sequential observations, and decide when sufficient evidence has been gathered to commit to a response. We consider the set of two-bar charts with tick values \{0, 1, 2, 3\} and latent bar heights $h_0,h_1$
taking single decimal place values at and between those tick values.
The agents are tasked with estimating the bar average rounded to one-decimal report grid; therefore, the task is not to reproduce inaccessible ground-truth bar values directly, but to produce a one-decimal report implied by the agent's internal estimate of the average. Despite this simple example, numerous model choices are still necessary in order to constrain the agents' cognitive approach to decision making. The Slow and Fast agents we present here are one plausible pair of such choices. We believe that our choices offer a testable proof-of-concept for dual-process decision making with realistic performance and interpretable parameters.

\subsection{Shared formalization and implementation choices}
\label{sec:shared_task}

Our design goals are inherited from Padilla~et~al.'s dual-process account of visualization-aided judgment~\cite{Padilla2018DecisionMaking}.
We build separate Fast and Slow agents, each producing action sequences justifiable as realistic strategies a human might use to complete the average-estimation task. The Slow model supports bar-wise reasoning while the Fast utilizes an "eye-balled" average estimation with reduced intermediate structure. In both cases, working memory and tick salience bias are represented as explicit computational variables.
In what follows, we write down explicit hidden states, observation models, working-memory dynamics, and report semantics for both analytic and heuristic judgment strategies for determining the average value of a two-bar chart.

\begin{figure}[t!]
    \centering
    \includegraphics[width=\linewidth]{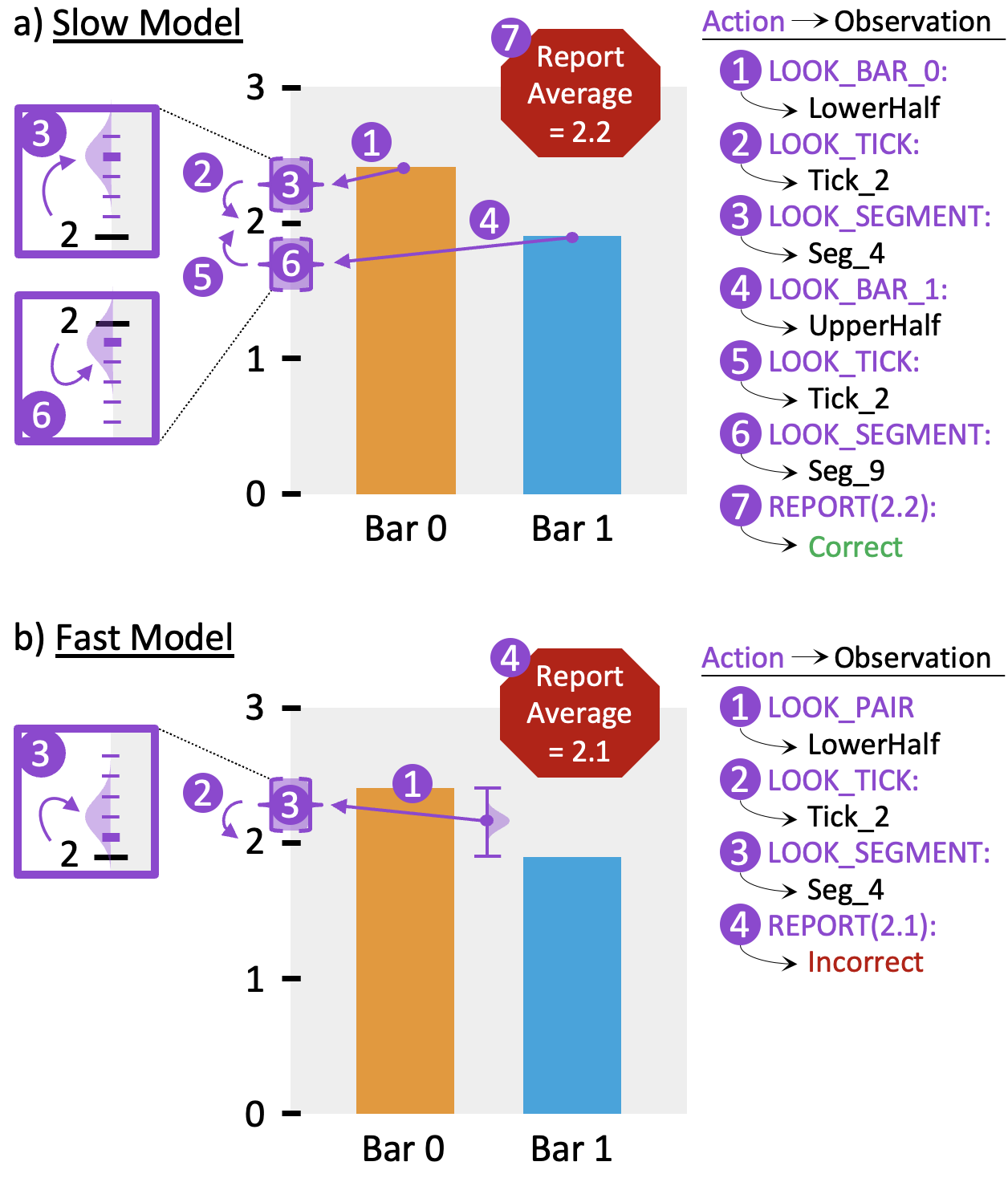}
    \vspace{-15pt}
    \caption{Diagram of the intended cognitive strategies used for design of the Slow and Fast models with actions and resulting observations.
    In panel (a), the Slow agent identifies the height of each bar and then reports the correct average. In panel (b), the Fast agent estimates the height of the visual midpoint between the bars and reports a slightly incorrect average.
    Purple arrows and inset illustrations represent the agents actions. Numbered circles index the time step of each action. 
    }
    \label{fig:model_diagram}
\end{figure}

\paragraph{Shared action vocabulary.}
Both models implement a staged perceptual routine in which the agent first samples a coarse cue, then anchors that cue to the axis, then refines it within a tick interval, and finally decides whether to report.
For the Slow model, the agent determines the hight of each bar by comparing their vertical position to the y-axis ticks. In the Fast model, the agent roughly estimates the midpoint of the two bars, and then uses that midpoint as the reference for determination of height and y-axis units. In either case, at each timestep the agent may select from four classes of actions:
\begin{enumerate}[leftmargin=*,itemsep=2pt,topsep=2pt]
\item \texttt{LOOK\_BAR\_i/LOOK\_PAIR}: Establish relative vertical position of point-of-interest (POI, either bar top or approximate midpoint of two bar tops) relative to y-axis ticks. This action constitutes the horizontal saccade of the human eye from the POI to the y-axis. 
\item \texttt{LOOK\_TICK}: Shift visual focus to closest tick, determining which tick interval the POI lies in. 
\item \texttt{LOOK\_SEGMENT}: Refine vertical position of POI to single decimal place interval between ticks, this comes with some noise associated with the absence of single-decimal reference ticks.  
\item \texttt{REPORT}: Report a believed average value for the two bars. 
\end{enumerate}
Figure \ref{fig:model_diagram} illustrates the intended series of actions that comprise the cognitive strategy of the Fast and Slow models. In what follows, we show how the chosen Slow and Fast models share a staged perceptual logic while specializing the hidden-state architecture to instantiate distinct cognitive strategies.

\paragraph{Observation vocabulary.}

Both models use the same observation vocabulary, with the agents observation at time t given by the tuple:
\begin{equation}
o_t
=
\bigl(
o_t^{\mathrm{rel}},
o_t^{\mathrm{axis}},
o_t^{\mathrm{seg}},
o_t^{\mathrm{fb}}
\bigr)
\label{eq:app_obs_tuple}
\end{equation}
where \texttt{LOOK\_BAR\_i/LOOK\_PAIR} yields a height observation relative to the nearest tick or tick interval,
\begin{equation}
    o_t^{\mathrm{rel}}
\in
\{\texttt{Null},\texttt{OnTick},\texttt{LowerHalf},\texttt{Midpoint},\texttt{UpperHalf}\}
\end{equation}
If the agent then chooses the action \texttt{LOOK\_TICK}, the value of the nearest tick is observed,
\begin{equation}
    o_t^{\mathrm{axis}}
\in
\{\texttt{Null}\}
\cup
\{\texttt{Tick}_k\}_{k=0}^{n_{\mathrm{ticks}}-1}
\cup
\{\texttt{LowerTick}_k\}_{k=0}^{n_{\mathrm{ticks}}-2}
\end{equation}
Choosing the action \texttt{LOOK\_SEGMENT} yields the nearest first decimal place with uncertainty corresponding to the lack of precise ticks, 
\begin{equation}
    o_t^{\mathrm{seg}}
\in
\{\texttt{Null}\}\cup\{\texttt{Seg}_j\}_{j=0}^{9}
\end{equation}
And finally, when the agent decides to report a believed average value $r$ with the action \texttt{REPORT}$(r)$, it receives feedback on its answer,
\begin{equation}
o_t^{\mathrm{fb}}
\in
\{\texttt{Null},\texttt{Incorrect},\texttt{Correct}\}
\end{equation}

This shared action-observation structure implements the staged perceptual logic of both models. A first action provides only coarse information about a POI relative to the y-axis. A second action anchors that POI to a specific tick or tick interval. A third action refines the estimate within the anchored interval.

\paragraph{Known-action approximation and policy library.}

Both models also use a discrete task register that tracks what is currently selected and whether the current estimate has been anchored to the axis. In the implemented models, the agent infers internal cognitive contents but does not need to infer which perceptual operation it has just selected. Action selection is therefore performed over a small hand-designed library of short policies rather than over all combinatorial action sequences. This approximation keeps planning tractable and also captures the idea that human observers bring a small repertoire of useful strategies to familiar chart-reading tasks. We designed our policy library to incentivize the agent towards sequences:

\vspace{-3pt}
\begin{equation}
\label{eq:design_strategy}
\scalebox{0.95}{$ 
\texttt{LOOK\_BAR} \rightarrow \texttt{LOOK\_TICK} \rightarrow \texttt{LOOK\_SEGMENT} \rightarrow \cdots \rightarrow \texttt{REPORT}
$}
\end{equation}


\noindent with the model repeating the initial action sequence (or a subset of it) if necessary for refinement before ending its inference with \texttt{REPORT}. 
In the final one-decimal implementations, we use a planning horizon of four steps that mix information gathering and report actions.

\paragraph{Generative process, memory decay, and perceptual bias.}
The environment contains the true bar heights and generates observations in response to actions. The agent, by contrast, has access only to sampled cues and updates beliefs under its own generative model. This distinction is important for our treatment of systematic perceptual bias: the environment may emit observations biased toward salient integer ticks, while the agent's likelihood model remains unaware. Memory decay is shared by both Fast and Slow models. For any memory state \(i\), the next memory state \(j\) is governed by
\begin{multline}
p(m_{t+1}=j\mid m_t=i)
=
\rho_{\mathrm{forget}}\,\mathbf{1}[j=0]
+
\rho_{\mathrm{mem}}\,\mathbf{1}[j=i]
\\+
\bigl(1-\rho_{\mathrm{mem}}-\rho_{\mathrm{forget}}\bigr)\,\psi(j;i)
\label{eq:memory_decay}
\end{multline}
where \(j=0\) denotes forgetting, \(\rho_{\mathrm{mem}}\) is the retention probability, and \(\psi(j;i)\) is a local diffusion kernel over nearby values.
This allows an estimate to be retained, forgotten, or blurred toward neighboring values.

To model systematic perceptual bias, we allow observations to shift toward the nearest salient integer tick on both the course \texttt{LOOK\_BAR\_i/LOOK\_PAIR} action or refined \texttt{LOOK\_SEGMENT} action. 
For a true y-axis value \(v\) and nearest tick \(\tau(v)\), the environment will return observations centered around
\begin{equation}
\tilde{v}
=
(1-\lambda_{\mathrm{tick}})v
+
\lambda_{\mathrm{tick}}\tau(v)
\label{eq:tick_bias_shift}
\end{equation}
where $\lambda_{\mathrm{tick}}$ controls the degree of tick-salience bias, taking values from 0 to 1.
Depending on if tick bias is being applied to the course initial cue or the refined segment action, $\lambda_{\mathrm{tick}}$ is implemented as either \texttt{course\_tick\_anchor\_bias} or \texttt{segment\_tick\_anchor\_bias}. 
Because these biased values are present in the generative \emph{process} but absent from the agent's generative \emph{model}, the agent is unaware of its own bias. This differs from unbiased Gaussian perceptual noise, over which the agent can in principle plan. In the Slow model, tick-salience bias affects one bar-specific estimate at a time. In the Fast model, it affects the compressed average estimate directly. We therefore expect the Fast model to be more susceptible to early anchoring errors, whereas the Slow model should be more susceptible to failures caused by memory load and multi-step integration.

\paragraph{Report semantics.}

Finally, both models report from their \emph{internal memory contents}, not directly from the inaccessible ground-truth bars.
This is important conceptually: the models are process accounts of human judgment, not direct classifiers.
In the one-decimal task, the epistemic term in Eq.~\eqref{eq:efe_general} is evaluated over uncertainty in the \emph{report distribution}, rather than over all hidden micro states, because the behaviorally relevant question is which answer should be given.

\subsection{Slow model: sequential bar-wise estimation}
\label{sec:slow_model}

The Slow model operationalizes Type~2-style processing by maintaining \emph{two explicit intermediate estimates}, one for each bar.
Its hidden state is
\begin{equation}
x_t^{\mathrm{slow}}
=
\big(
r_t,\,
m_t^{(0)},\,
m_t^{(1)}
\big),
\end{equation}
where
\(r_t\in\{\texttt{UNSET},\texttt{BAR0},\texttt{BAR1},\texttt{BAR0\_ANCHORED},\texttt{BAR1\_ANCHORED}\}\)
is the task register and \(m_t^{(0)},m_t^{(1)}\) are working-memory states for the two bars. The Slow model is designed to follow the strategy of Eq.~\ref{eq:design_strategy} applied first to one bar and then to the other. The first action writes a coarse estimate for bar \(i\) into memory. The second anchors that estimate to the appropriate axis interval.
The third refines the estimate within that interval. As the two bars are represented separately, the model can in principle revisit one bar without overwriting the other.

The final report is produced by projecting the joint memory state onto the one-decimal report grid.
If both bar memories are set, the implied report value is the rounded mean of the two remembered bar values.
If one or both memories are unset, the projected report distribution remains broad.
Writing \(q_t^{\mathrm{slow}}\) for the posterior over the full hidden state, the induced report distribution is
$q_t^{\mathrm{rep}}
=
P_{\mathrm{slow}}(q_t^{\mathrm{bar0}}, q_t^{\mathrm{bar1}})$,
where \(P_{\mathrm{slow}}\) deterministically maps each joint memory state to the corresponding one-decimal report value.
We need this projection to translate the model's stored bar values into the required rounded average.

The Slow model is therefore structurally aligned with a more deliberate and decomposed judgment process.
Its main advantage is that it preserves interpretable intermediate structure.
Its main vulnerability is that this structure must be maintained across multiple steps.
Even when each local perceptual update is reasonable, the final report can be wrong because one bar-specific intermediate representation has decayed, diffused, or remained insufficiently resolved before report.
This gives the Slow model a clear Type~2 failure mode: errors caused by the fragility of sequential integration and working memory.

\subsection{Fast model: compressed average-first estimation}
\label{sec:fast_model}

The Fast model operationalizes Type~1-style processing by maintaining a \emph{single compressed estimate} of the average directly, rather than separate estimates for the two bars.
Its hidden state is
\begin{equation}
x_t^{\mathrm{fast}}
=
\big(
r_t,\,
m_t^{(\mathrm{avg})}
\big),
\end{equation}
where
\(r_t\in\{\texttt{UNSET},\texttt{PAIR},\texttt{PAIR\_ANCHORED}\}\)
is a compact task register and \(m_t^{(\mathrm{avg})}\) is a memory state defined directly on the one-decimal report grid \(\mathcal{R}\). Here the first action does not separately estimate the two bars.
Instead, it acquires a coarse cue about the \emph{pairwise average} itself.
The axis and segment actions then anchor and refine that compressed average estimate.
Because the memory variable already lives on the report grid, the report projection is simpler than in the Slow model:
$q_t^{\mathrm{rep}}
=
q_t^{\mathrm{avg}}$,
and, for set memory states, \(P_{\mathrm{fast}}\) is effectively an identity map from average memory to report value.

The Fast model uses the same general memory-decay law as Eq.~\eqref{eq:memory_decay}, but its main limitation is not memory structure; it is \emph{representational compression}.
By encoding only a single average estimate, the model has fewer internal degrees of freedom available for later correction.
This makes it a natural model of fast, gist-based judgment: fewer internal variables, fewer integration steps, and earlier commitment.

\section{Results}
\label{sec:results}

\subsection{Baseline model validation: }
\label{sec:baseline_validation}

Our evaluation proceeds in two stages: first, characterizing the general behavioral and performance differences between the Fast and Slow agents under ideal conditions, and second, isolating their respective vulnerabilities to failure modes. The following experiments are hypothesis-driven stress tests of two idealized task strategies, not independent evidence that these agents discover distinct cognitive failures in humans. Because the Fast and Slow agents are constructed around different representational commitments, the purpose of these results is to quantify the inspectable failure modes implied by those commitments.

To establish the baseline, we initialized both models with high memory-retention and no tick-bias. Even under these ideal conditions, we expect the Fast model to exhibit a naturally lower accuracy due to its compressed, heuristic cognitive strategy. After establishing these baseline dynamics, we perturb the models by varying a single mechanistic parameter associated with each hypothesized failure mode: tick anchoring in the Fast model and memory decay in the Slow model.

The selected parameter values for this initial analysis of model performance are displayed in the appendix in Table \ref{tab:appendix_defaults}. The only difference in parameters values used for the Fast and Slow models are the noise width on the initial \texttt{LOOK\_BAR\_i/LOOK\_PAIR} used for course interval/tick alignment. Observations for the Fast model were given a standard deviation of $\texttt{pair\_obs\_sigma} = 0.22$ while the Slow model received $\texttt{bar\_obs\_sigma} = 0.0$. This is intended to model the imprecision of the "eye-balled" midpoint between the two bars, compared to the visually salient bar-tops used by the Slow model. When either agent hits their time deadline (shorter for the Fast agent than the Slow), it is forced to make its best guess at the average.

\begin{figure}[!t]
    \centering
    \includegraphics[width=1.0\linewidth]{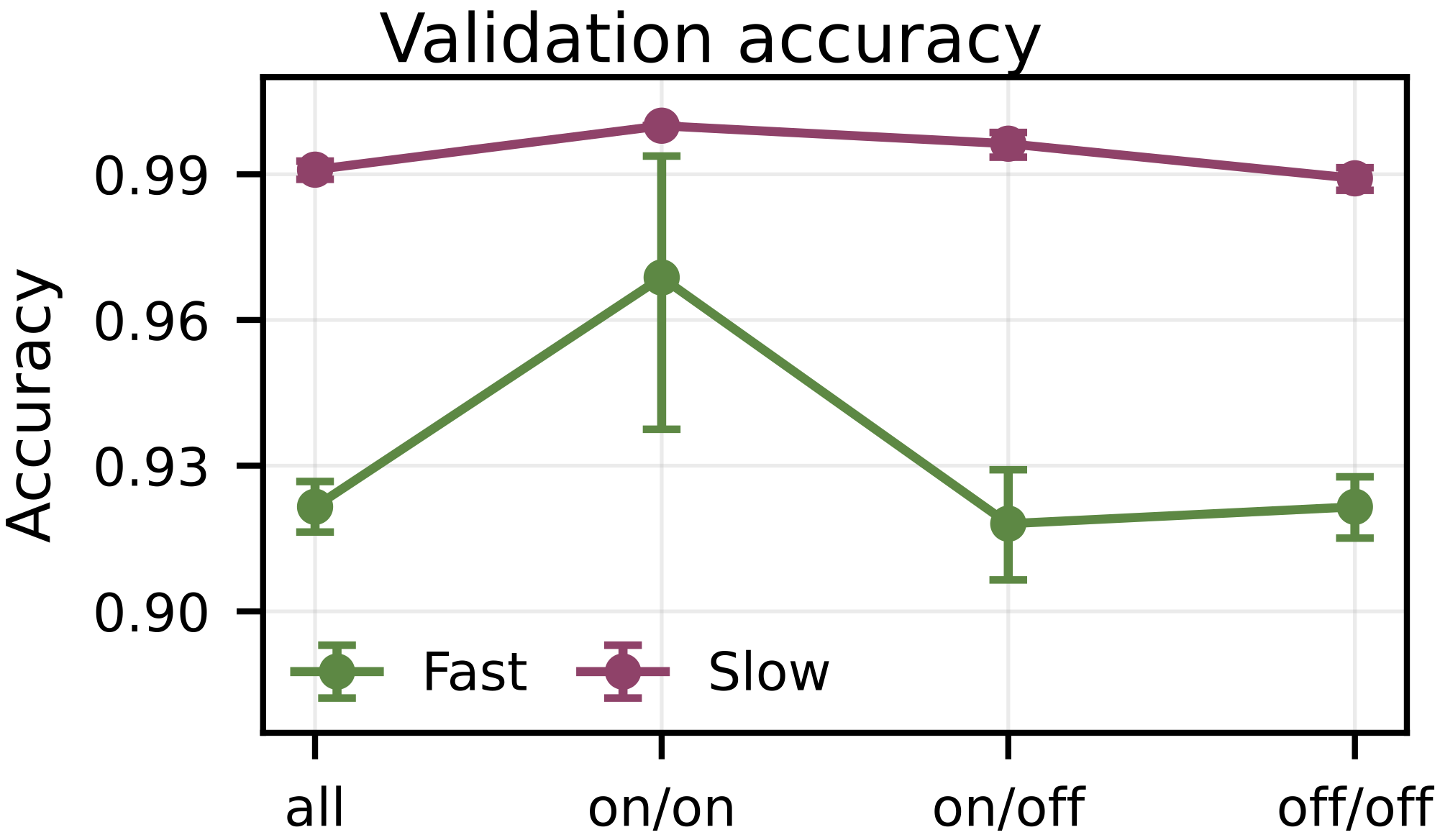}
    \vspace{-15pt}
    \caption{
    Baseline validation of Fast and Slow models in determining the average of a two-bar chart across all bar values between 0.0 and 3.0. Agents ran on unique combination of bar heights across 10 random seeds. Fast and Slow accuracy rates are plotted for both bars on-tick (on/on), one bar on-tick (on/off), and both bars off-tick (off/off).
    }
    \label{fig:baseline_validation}
\end{figure}

For our baseline validation, the Fast and Slow agents were run on every possible combination of bar height values across the 0.0 -- 3.0 range, each with 10 random seeds, making for a total of 9610 runs of each model. Figure \ref{fig:baseline_validation} displays the accuracy of reported bar averages. Panel A displays overall accuracy, and then category specific accuracy split between bar-heights both on ticks (on/on), one bar on a tick and another off-tick (on/off), and both bars off-tick (off/off). Across all categories, the Fast model fails more often, with an overall accuracy of 0.92. The Slow model rarely failed, with overall accuracy of 0.98. 

To assess trends in which bar-heights the Fast model failed to determine the average bar value, Figure \ref{fig:baseline_validation_2} maps out the difference in failure rate between the two models for each unique combination of bar heights. A diagonal banded structure is apparent, indicating that when the two-bar average is 1.0 or 2.0, the Fast model performs better. This is consistent with the visual salience of the integer ticks.  

We note here that the task specific failure rates examined in this baseline validation are clear metrics to parameterize these models from user studies. This point of comparison to human users would provide a test of the basic cognitive strategies chosen for our agents. We continue this discussion in the Section \ref{sec:future}. Also, a primary advantage of Active Inference is the interpretable timeseries of posterior belief distributions output by the agents. Appendix \ref{sec:heuristic_fail_ex} includes an example set of cognitive traces for a case where the Fast model fails and the Slow model succeeds. In that example, we can determine from the posterior beliefs that the Fast model mistakenly committed after an erroneous result from the noisy \texttt{LOOK\_SEGMENT} action. This type of causal failure is a prime candidate for user study validation or falsification.

\subsection{Type 2 Failure: Memory Decay}

To quantify the consequences of memory dependence in the two strategies,
we exposed both agents to increasing memory decay in the bar-chart interpretation task. This perturbation is expected to disproportionately affect the Slow strategy because it stores separate intermediate estimates for each bar before combining them into an average. The experiment therefore serves as a mechanistic stress test: it measures how strongly the sequential analytic strategy depends on maintaining intermediate beliefs across multiple perceptual actions.

As described in Eq.\ \ref{eq:memory_decay}, memory decay was implemented as diffusion in the agents posterior beliefs. Both agents were parameterized by the probability of memory retention \texttt{mem\_retention} ($\rho_\mathrm{mem}$) and the probability of forgetting entirely \texttt{mem\_forget} ($\rho_\mathrm{forget}$). The remaining probability can be interpreted as the memory decay rate.   

\begin{figure}[!t]
    \centering
    \includegraphics[width=\columnwidth]{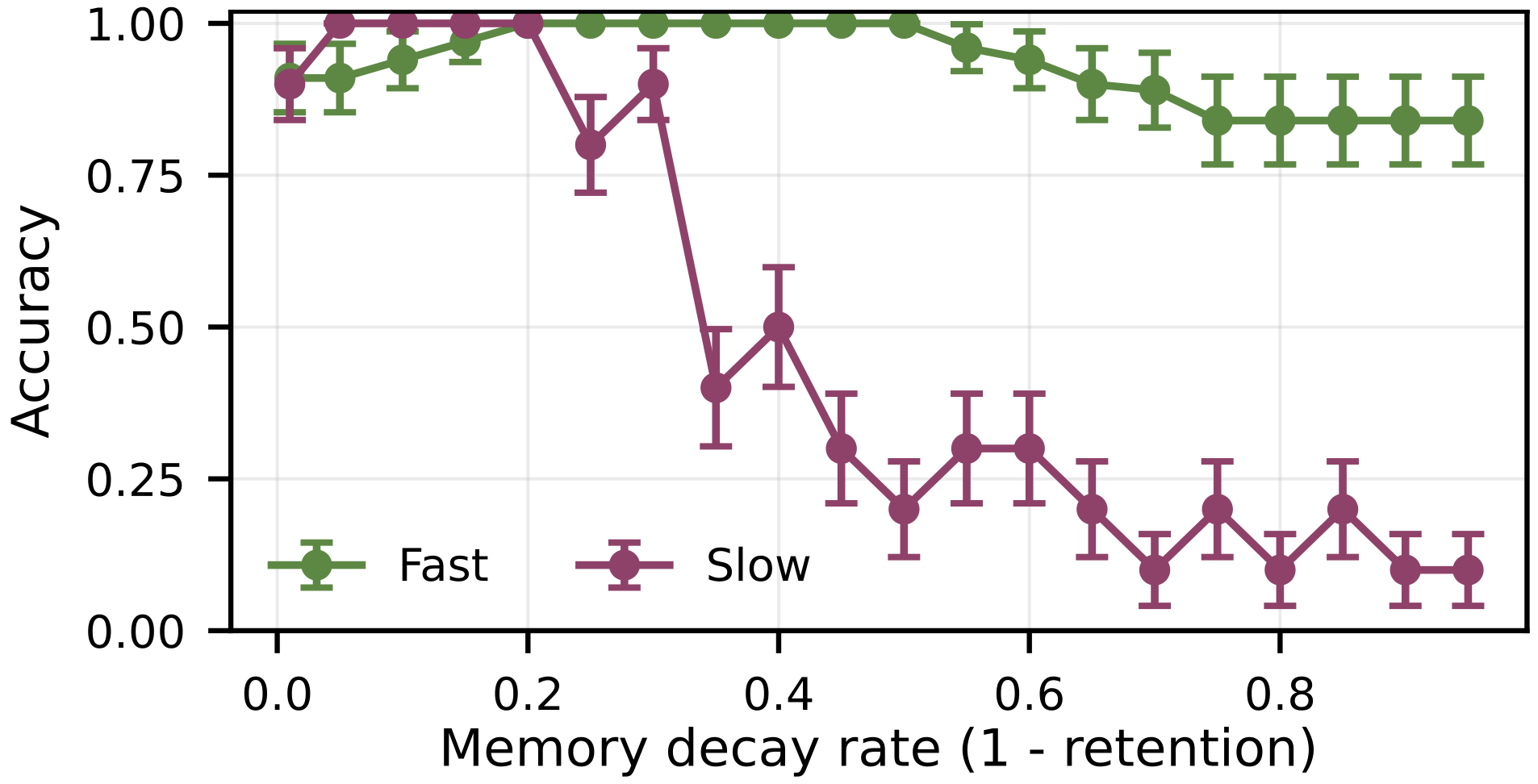}
    \vspace{-15pt}
    \caption{Both Fast and Slow models were run with increasing memory decay on 10 random pairs of bar heights across 10 random seeds. Panel A plots accuracies computed per pair of bar heights (across random seems) and then averaged to generate data points with one standard deviation plotted. Accuracy for both model rises due to increased belief uncertainty leading to deliberation. As the memory decay rate increases beyond 0.2, Slow model accuracy falls sharply due to increased time required for the its cognitive strategy. 
    See Fig.~\ref{fig:memory_steps} for timestep data.
    }
    \label{fig:memory}
    \vspace{-10pt}
\end{figure}

The fraction of memory not retained or forgotten (mass shifted to \texttt{UNSET} state) diffuses to neighboring memory states following a discrete Gaussian distribution with width \texttt{mem\_sigma}. For the memory decay experiment, we set \texttt{mem\_sigma} $= 5.0$. With this parameter and the shared values described in Table \ref{tab:appendix_defaults}, ran both models sweeping across \texttt{mem\_retention} 20 values from 0 to 1, each computed on 10 bar height combinations each with 10 random seeds (100 model runs per data point). Figure \ref{fig:memory} visualizes the results of this experiment, showing the stark drop in Slow model accuracy with increasing memory decay. Interestingly, both models improve in accuracy before declining, reaching equally high-accuracy at a decay rate ($1 -$ \texttt{mem\_retention}) of 0.2. Upon examination of the mean number of time steps (see 
Fig.\ \ref{fig:memory_steps}) and output cognitive traces, its clear that a small amount of memory decay encourages more deliberation. Since memory decay is applied to posterior beliefs, the agents become increasingly less certain as they take actions at each time step. The drive for epistemic certainty leads to increased time spent trying to refine beliefs. This increased deliberation benefits the Fast model more than the slow model, since with negligible memory decay the Fast model tended to commit too 
early. 

\subsection{Type 1 Failure: Tick-Salience Bias}

To quantify the consequences of representational compression, we exposed both agents to increasing tick-salience bias.
This perturbation is expected to disproportionately affect the Fast model because it estimates the bar-pair average through a compressed visual representation of the bar average. 
By contrast, the Slow model estimates the two bar heights separately, so anchoring errors can partially cancel when the separate estimates are averaged. This experiment is not to show that the Fast model unexpectedly develops anchoring bias, but to measure the quantitative signature implied by the compressed strategy.

\begin{figure}[!t]
    \centering
    \includegraphics[width=\linewidth]{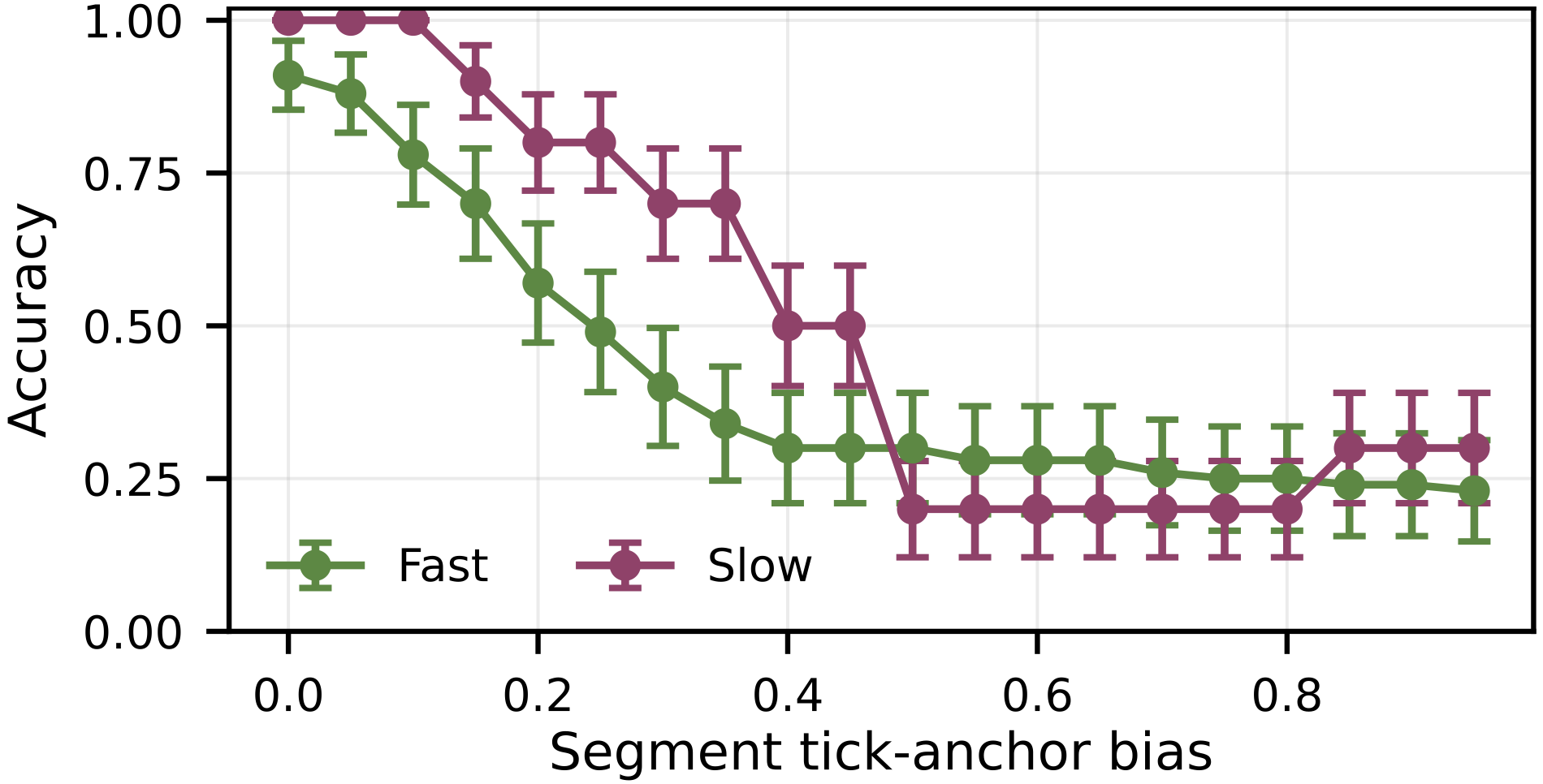}
    \vspace{-20pt}
    \caption{Both Fast and Slow models were run with increasing tick salience bias on the \texttt{Segment\_i} observation following a \texttt{LOOK\_SEGMENT} action. The tick bias parameter \texttt{segment\_tick\_bias} shifts the centroid of the noisy observation within the environment but not withing the agent. Both agents were run on 10 random pairs of bar heights across 10 random seeds. Accuracies are computed per pair of bar heights (across random seems) and then averaged to generate data points with one standard deviation. Accuracy for both model falls as tick bias increases, but the Fast model accuracy decays more rapidly. 
    See Fig.~\ref{fig:tick_steps} for timestep data.}
    \label{fig:tick_bias}
    \vspace{0pt}
\end{figure}

Figure \ref{fig:tick_bias} visualizes the accuracy of both models under increasing value of \texttt{segment\_tick\_anchor\_bias}, the parameter controlling tick bias applied to the \texttt{LOOK\_SEGMENT} actions. For this experiment, memory effects were turned off. Tick bias is encoded in the environment but not the generative models. As \texttt{segment\_tick\_anchor\_bias} increases, we see the accuracy of both models fall, but the fast model succumbs more quickly. While Fast accuracy diminishes imminently, Slow accuracy does not begin to fall until \texttt{segment\_tick\_anchor\_bias} $= 0.15$. The rate of accuracy decay is also less for the Slow model than the Fast. This is until \texttt{segment\_tick\_anchor\_bias} $= 0.5$, where accuracy between the two models is comparable. The existence of parameter regimes yielding equivalent model accuracies highlights the need for future hypothesis testing on empirical data to consider fitting to additional model outputs such as failure maps, fixation sequences, step counts, and belief trajectories. 

\section{Discussion}
\label{sec:discussion}

The main contribution of this work is not simply that Active Inference can be made to read a bar chart. 
Instead, we demonstrate that qualitative cognitive strategies for visualization use can be translated into executable models with explicit internal states, explicit failure modes, and quantitative behavioral predictions. 
In our implementation, the Fast and Slow agents are mechanistic hypotheses about how a viewer might solve the task: one by maintaining a compressed estimate of the average directly, the other by establishing separate estimates for the height of each bar. 
Because these assumptions are written down explicitly, disagreement with human behavior would not merely indicate poor predictive performance; it would indicate that the chosen cognitive strategy or its parameterization is wrong. In this sense, the value of the models is not only descriptive realism, but quantitative falsifiability. 

The results support this interpretation by showing the two architectures fail in systematically different ways. At baseline, the Fast model achieved an overall accuracy of 0.92, while the Slow model achieved 0.98 across the full sweep of one-decimal bar pairs, and the pair-specific failure map revealed scenarios where the two strategies diverged and aligned (Fig.~\ref{fig:baseline_validation}). 
Specific cases where model performance differs can be examined mechanistically.  
The example cognitive trace for the case of bar heights 2.4 and 1.9 reveals the Fast model committing to an incorrect average after a single mistake in refinement (Fig.~\ref{fig:heuristic_failure_trace}).
%
The parameter sweep experiments further exemplify testable differences between the two cognitive strategies. Increasing memory decay disproportionately harms the Slow model, with accuracy falling sharply once decay becomes substantial (Fig.~\ref{fig:memory}), while increasing tick-salience bias degrades both agents but causes the Fast model to fail more rapidly (Fig.~\ref{fig:tick_bias}). 
These selective vulnerabilities are not surprising post hoc discoveries; they are the expected consequences of the strategy definitions expressed as quantitative predictions: 
accuracy differences, stimulus-specific error structure, step-count differences, and selective sensitivity to perturbations of memory and perceptual bias. 

For visualization research, the significance of this framework is that it predicts more than aggregate success or failure. It predicts where in stimulus space each strategy fails, how long it takes before reporting, and how those outcomes change as interpretable parameters are varied. 
This is what makes the models useful for user studies. The baseline failure-rate maps, the memory-decay curves, and the bias-sensitivity curves are all quantitative signatures that could be used to parameterize or falsify the proposed cognitive strategies against human data. 
If user studies were designed to 
test these predictions,
we could test the hypothesized decomposition of Fast and Slow judgment, iterate on model factorization, and eventually reveal indicators for when viewers rely on compressed heuristic versus sequential analytic strategies. Until verified as accurate human stand-ins, these models remain tentative mechanisms utilized exclusively to test cognitive hypotheses.

\section{Limitations and Future Work}
\label{sec:future}

The present work is a proof of concept for a modeling methodology, not a complete account of human visualization use. The task is deliberately narrow, the observation model is symbolic rather than pixel-based, and the policy library is hand-designed. These simplifications were useful for isolating the cognitive layer of the problem and for keeping the resulting traces interpretable, but they also limit the behavioral richness of the current agents. More realistic tasks will require richer visual scenes, more flexible policies, and perceptual front-ends that propagate uncertainty from the display itself into the generative model. A limitation of our dual-process implementation is that the present Fast and Slow agents are only one pair of formalizations among many that could instantiate dual-process ideas. The Slow model uses bar-specific intermediate beliefs, while the Fast model uses a single compressed average estimate, but these are modeling choices rather than evidence based decompositions of human interpretation of bar charts. Their scientific value is that they demonstrate distinct measurable signatures: different baseline error structure, different step counts, stronger Slow sensitivity to memory decay, and stronger Fast sensitivity to tick-salience bias. 

The most important next step is therefore empirical. Because the models produce stimulus-conditional predictions rather than only aggregate accuracy, they can be tested against several kinds of human data at once: overall accuracy, error magnitude, reaction times or fixation counts, and full maps of failure rates across bar pairs. Eye-tracking data will be especially valuable, because the Slow model predicts more extended sequential sampling and greater dependence on maintaining bar-specific information, whereas the Fast model predicts earlier commitment and greater susceptibility to salient anchors. If human viewers do not exhibit these predicted task failures, then the strategies encoded by the present models should be revised or rejected. In that sense, future work is not simply about scaling the models up, but about confronting explicit mechanistic hypotheses with quantitative behavioral evidence.

This empirical perspective also suggests a concrete path for generalizing beyond the present two-bar average-estimation task.
Modeling other visualization-interpretation will require identifying mechanistically defendable hidden states, observation modalities, actions, preferences, and policy structure.
Predictive scan-path models and ``virtual viewer'' systems could help address one of the central model-design challenges in Active Inference: specifying the observation and action spaces for new chart-reading tasks. 
In the present proof of concept, the observation modalities and policy library are hand engineered from a hypothesized strategy for bar-chart average estimation. 
This is appropriate for a controlled demonstration, but leaves open how future models should be constructed for tasks such as value retrieval, comparison, filtering, aggregation, or extrema detection \cite{brumar2026typologydecisionmakingtasksvisualization}. 
Future work could use measured or predicted task-conditioned scan-paths as an empirical scaffold for this step, identifying which chart elements viewers sample during different tasks and translating those sampled elements into candidate observation modalities, such as mark-value, axis-anchor, legend, label, comparison, or contextual-task observations. 
Recurrent fixation transitions could likewise suggest useful policy templates, such as mark-to-axis routines for value retrieval, mark-to-mark routines for comparison, or sequential inspection routines for filtering and extrema detection. 
Thus, scan-path models would not serve only as validation targets; they could act as empirical proposal mechanisms for constructing task-specific generative models. 
Active Inference would then provide the complementary mechanistic account of how those observations update beliefs, how cognitive biases and limitations affect the chain of reasoning, and when\/why the viewer responds. 

Acknowledging current limitations, our hope is that this work leads to operable mechanistic models of human cognition. Another limitation that must be addressed is that in our proof-of-concept the Fast and Slow agents are evaluated as independent, parallel formalizations. However, human cognition does not run two separate brains; it dynamically shifts between heuristic and analytic processing based on internal confidence and task demands. A true computational formalization of Padilla et al.’s framework \cite{Padilla2018DecisionMaking}, requires a unified, hierarchical Active Inference architecture. In future work, we propose building a meta-cognitive ``arbiter''. Operating at a slower timescale, this higher-order agent would monitor the expected free energy ($G$) of the Type 1 subroutine. If the visual encoding is ambiguous, causing the epistemic uncertainty to exceed a certain threshold, the arbiter would autonomously trigger a cognitive switch, shifting the agent into the sequential, working-memory-intensive Type 2 policy space. This hierarchical formulation would allow designers to 
estimate candidate visual thresholds that may correspond to increased cognitive effort. While our forward simulations yield hypothesized human-like failure modes, a critical next step is the inverse problem: fitting these models to empirical data. Mapping noisy behavioral traces (e.g., eye-tracking, reaction times) onto discrete POMDP matrices is a non-trivial challenge requiring techniques like Inverse Active Inference and Bayesian Model Reduction. For instance, empirical gaze-transitions could parameterize policy priors, while contrast sensitivity metrics could fit the observation model ($A$-matrix). Developing this parameter-fitting pipeline is an important step toward empirically grounded, predictive visualization evaluation.

Finally, we must address the role of mechanistic cognitive models alongside the rapid rise of Large Language Models (LLMs). It is tempting to suggest that simulating user behavior could be solved simply by prompting an LLM to evaluate a chart. However, LLMs are fundamentally statistical sequence predictors; they do not possess a grounded "world model" of human perception, nor do they natively experience working memory constraints, visual saccade costs, or cognitive fatigue. They cannot be reliably parameterized to fit a specific user's eye-tracking trace. Active Inference provides exactly what LLMs lack: a causal, inspectable, and mathematically rigorous ledger of the decision-making process. Rather than viewing these as competing paradigms, future work should explore their intersection, leveraging the LLM as an intuitive interface and orchestrator around the detailed cognitive agent engine. An LLM could act as a sophisticated wrapper around the agent engine, providing the interface for human users to interact with and queue up visualization studies and analyze specific visual encodings. The LLM can then run that agent engine for the specified encodings and provide back the `agent study data', effectively transforming mechanistic modeling into an accessible and powerful analytical tool for visualization research.

\section{Conclusion}
\label{sec:conclusion}

The evaluation of visualizations has 
long been grounded in empirical evaluation. 
Such studies are essential for identifying
\emph{when} a design 
succeeds or fails, but they do not by themselves always provide an executable account of \emph{why} those outcomes arise.
In this paper, we introduced a mechanistic approach using Active Inference. By framing chart interpretation as a unified process of observation and action, we translated qualitative cognitive theory into an executable and simulatable problem. 
By framing bar-chart average estimation as sequential inference and control, we 
complement empirical accounts of visualization success and failure with an executable process model for mechanistic hypothesis testing.
Active Inference enables explicit modeling of how a viewer might gather information, maintain intermediate beliefs, and commit to a response. Rather than building a single agent that was optimized for accuracy, we constructed matched Type 1 and Type 2 agents that explicitly model the architectural trade-offs of human cognition. The value of this approach lies in its ability to generate 
explicit, testable
failure modes. Our results demonstrated that the Fast agent's compressed representation made it distinctly vulnerable to perceptual anchoring (i.e., tick-salience bias), while the Slow agent's reliance on sequential integration made it highly sensitive to working-memory decay. 
By isolating these variables, we 
illustrated how visualization interpretation errors can arise as structured consequences of cognitive strategy interacting with visual design.We do not propose that synthetic agents should replace human participants in visualization research. Instead, we offer this Active Inference framework as a mechanistic scaffold and call to action. To realize the vision of \emph{generative visualization design}, where designers can computationally simulate, evaluate, and debug visual encodings 
\emph{in silico} 
before or alongside human-subject studies,
future work should treat human behavioral data not only as evidence for evaluating designs, but also as evidence for parameterizing, testing, and falsifying mechanistic models.
By fitting future Active Inference models to rich empirical traces such as eye tracking and reaction times, we can move beyond simply describing visualization performance, and 
begin generating causal, testable predictions about it.

\section*{Supplemental Materials}
\label{sec:supplemental_materials}

Supplemental materials include attached appendices and software available at <removed for review>. 
\url{https://github.com/NatLabRockies/AIF_for_Vis}. 
In Appendix~\ref{sec:additional_results}, we include more comprehensive analysis from the numerical experiments conducted in Section~\ref{sec:results}.

Appendix~\ref{app:formalism} detailed mathematics of the active inference formalism used in this work. Appendix~\ref{app:shared_task} contains implementation details  shared by the two models. Appendix~\ref{app:architectures} gives the full technical factorization of the Slow and Fast architectures. 

\section*{Figure Credits and Copyrights}
\label{sec:figure_credits}

Figure \ref{fig:teaser} was produced by prompting Gemini 3 Pro Image with author generated illustrations. 
All other figures are original work of the authors, including Fig.~\ref{fig:padilla} being a reinterpretation from Ref.~\cite{Padilla2018DecisionMaking}.

\section*{Acknowledgments}
This work was authored by the National Laboratory of the Rockies for the U.S. Department of Energy (DOE), operated under Contract No. DE-AC36-08GO28308. Funding provided by Laboratory Directed Research and Development (LDRD). The views expressed in the article do not necessarily represent the views of the DOE or the U.S. Government. The U.S. Government retains and the publisher, by accepting the article for publication, acknowledges that the U.S. Government retains a nonexclusive, paid-up, irrevocable, worldwide license to publish or reproduce the published form of this work, or allow others to do so, for U.S. Government purposes. We thank project collaborators for discussions on Active Inference and visualization cognition.

\bibliographystyle{abbrv-doi-hyperref-narrow}
\bibliography{refs}

\appendix

\section{Additional Results}
\label{sec:additional_results}

\subsection{Baseline Validation: Additional Details}
\label{app:baseline_additio }

To make the reported experiments reproducible, Table~\ref{tab:appendix_defaults} records the default settings used in the baseline validation experiments.

\begin{table}[h]
\centering
\caption{Default comparison settings used in the main validation scripts. These are experimental defaults for the reported Fast/Slow comparisons, not intrinsic parts of the architectural definition.}
\label{tab:appendix_defaults}
\begin{tabular}{ll}
\toprule
Setting & Value \\
\midrule
\(n_{\mathrm{ticks}}\) & \(4\) \\
\(\Delta_h\) (\texttt{height\_step}) & \(0.1\) \\
policy horizon (\texttt{policy\_len}) & \(4\) \\
policy precision \(\gamma\) & \(8.0\) \\
policy evaluation mode & \texttt{rollout} \\
epistemic target & \texttt{report} \\
report action instant & \texttt{True} \\
action selection & deterministic \\
\(\rho_{\mathrm{mem}}\) (\texttt{mem\_retention}) & \(0.97\) \\
\(\rho_{\mathrm{forget}}\) (\texttt{mem\_forget}) & \(0.005\) \\
\(\sigma_{\mathrm{mem}}\) (\texttt{mem\_sigma}) & \(0.25\) \\
non-report cost \(c_{\mathrm{nr}}\) & \(0.35\) \\
segment cue width (\texttt{segment\_sigma}) & \(0.045\) \\
coarse tick bias (\texttt{course\_tick\_anchor\_bias}) & \(0.0\) \\
segment tick bias (\texttt{segment\_tick\_anchor\_bias}) & \(0.0\) \\
\midrule
Fast-only coarse cue width (\texttt{pair\_obs\_sigma}) & \(0.22\) \\
Fast-only trial horizon \(T\) & \(7\) \\
\midrule
Slow-only coarse cue width (\texttt{bar\_obs\_sigma}) & \(0.00\) \\
Slow-only trial horizon \(T\) & \(10\) \\
\bottomrule
\end{tabular}
\end{table}

\begin{figure}[!t]
    \centering
    \includegraphics[width=1.0\linewidth]{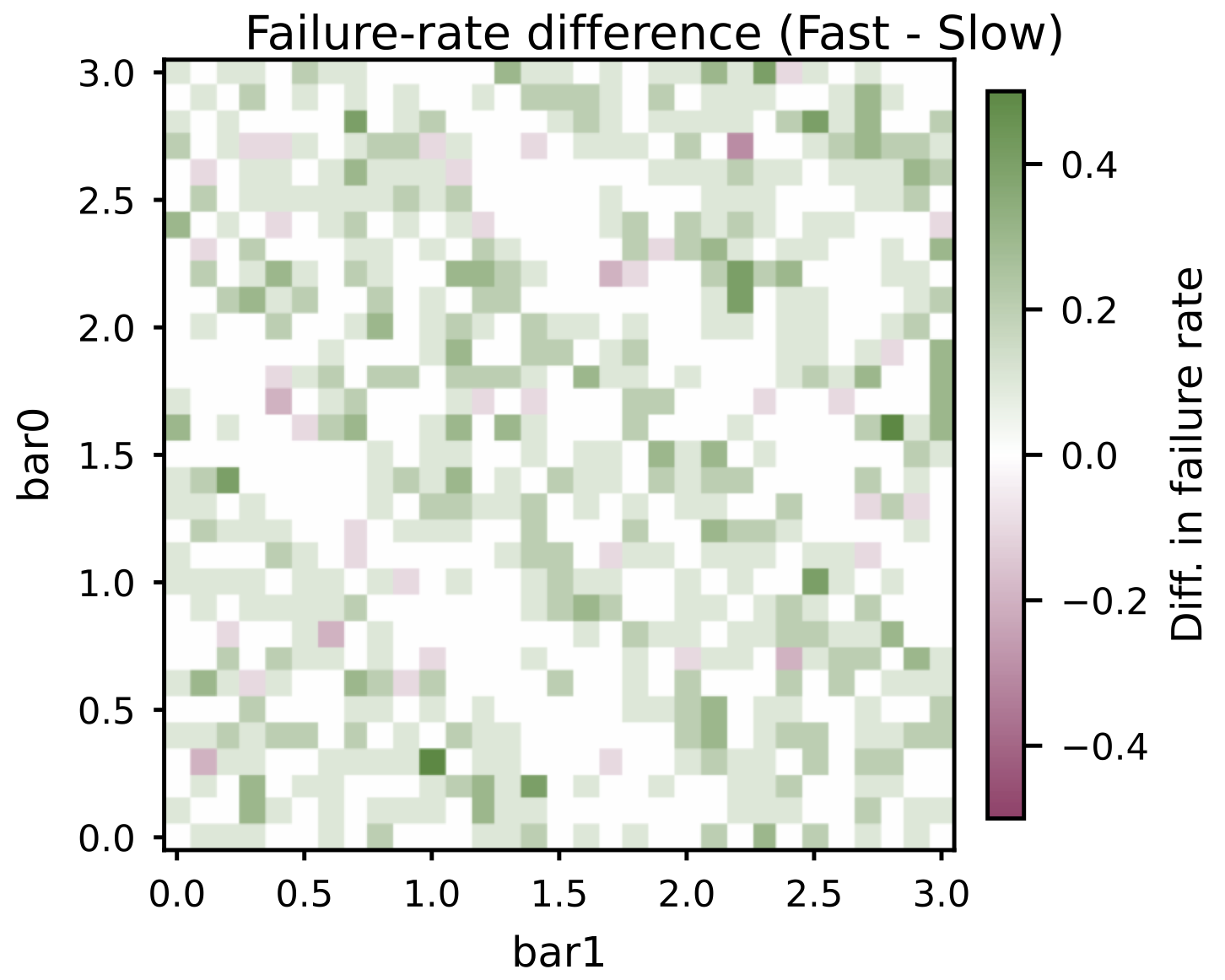}
    \caption{
    Baseline validation of Fast and Slow models for task of determining the average of a two-bar bar chart across all single decimal place bar values between 0.0 and 3.0. Agents were run on each unique combination of bar heights across 10 random seeds.
    Accuracy for each unique combination of bar heights is plotted in a grid, visualizing the difference in Fast and Slow model accuracy. Green regions indicate Fast model failure. Erratic purple regions denote slow model failure. 
    }
    \label{fig:baseline_validation_2}
\end{figure}

To assess trends in which bar-heights the Fast model failed to determine the average bar value, Figure \ref{fig:baseline_validation_2} maps the difference in failure rate between the two models for each unique combination of bar heights. Red areas indicate bar pairs for which the Fast agent failed to determine the correct average more often (across 10 random seeds) than the Slow model. A diagonal banded structure is apparent, indicating that when the two-bar average is 1.0 or 2.0, the Fast model performs better. This is consistent with the visual salience of the integer ticks.  

Fig.~\ref{fig:sup_validation} visualizes comprehensive statistics on the baseline validation over all possible bar-height combinations. The main validation script instantiates the Fast agent and the Slow agent to simulate chart reading over exhaustive sets of bar pairs, and summarize behavior in terms of exact-match accuracy, report rate, mean absolute error, mean steps, and pair-specific failure structure.

\subsection{Baseline Inaccuracy of Fast Model: Heuristic Error.} \label{sec:heuristic_fail_ex}

\begin{figure*}[t!]
    \centering
    \includegraphics[width=\linewidth]{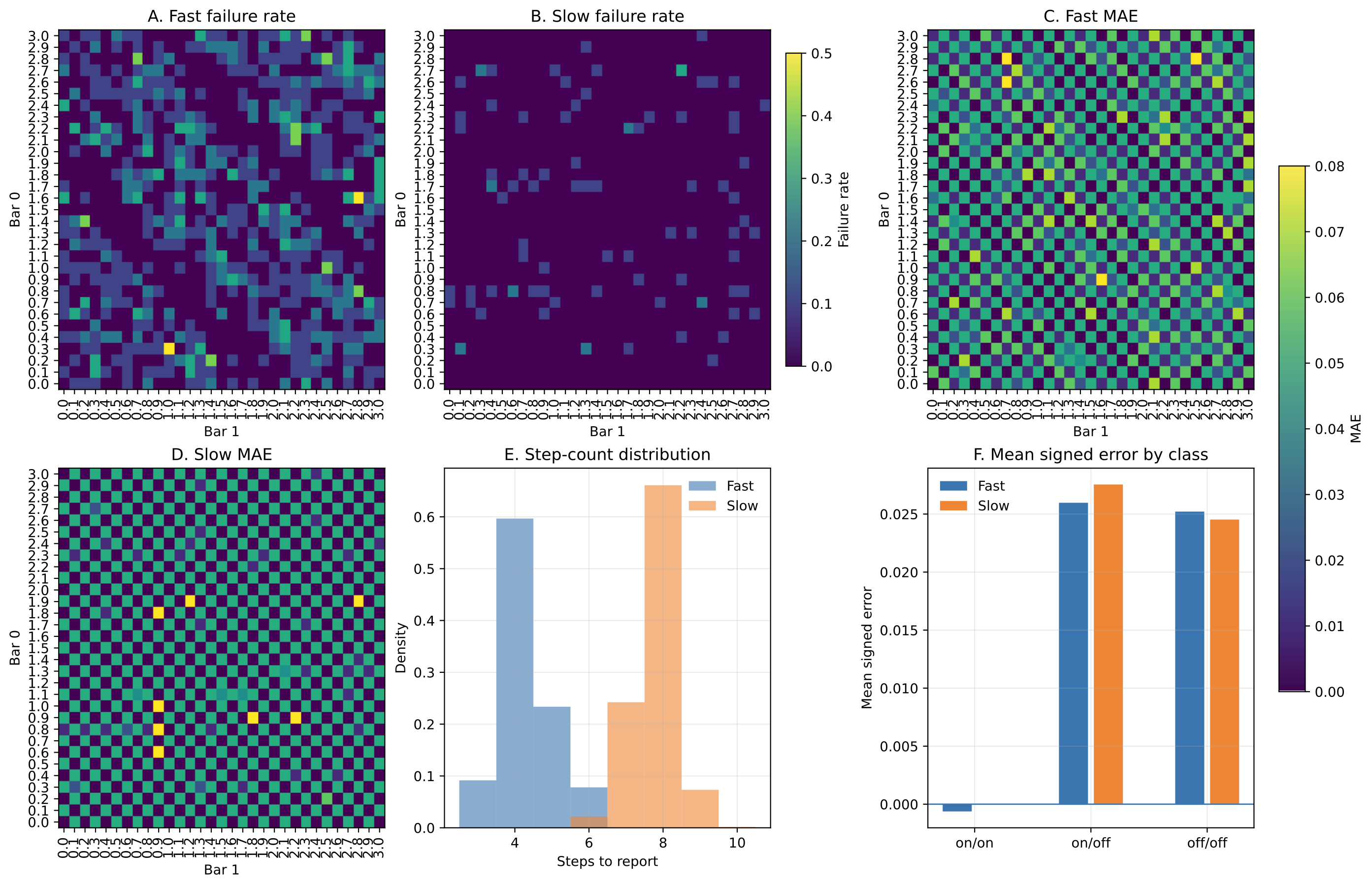}
    \caption{Comprehensive statistics on the baseline validation experiment presented in Section \ref{sec:baseline_validation}.
    }
    \label{fig:sup_validation}
\end{figure*}

When examining the cases of failure for the Fast model in baseline sweep across all combinations of bar values (Fig.~\ref{fig:sup_validation}), we notice that the mean error is small. Examining the cognitive traces of the model in these cases reveals that the heuristic structure of the Fast model (relying on the noisy midpoint between bar tops) occasionally forces it into premature confidence. 

\begin{figure}[t!]
    \centering
    \includegraphics[width=\linewidth]{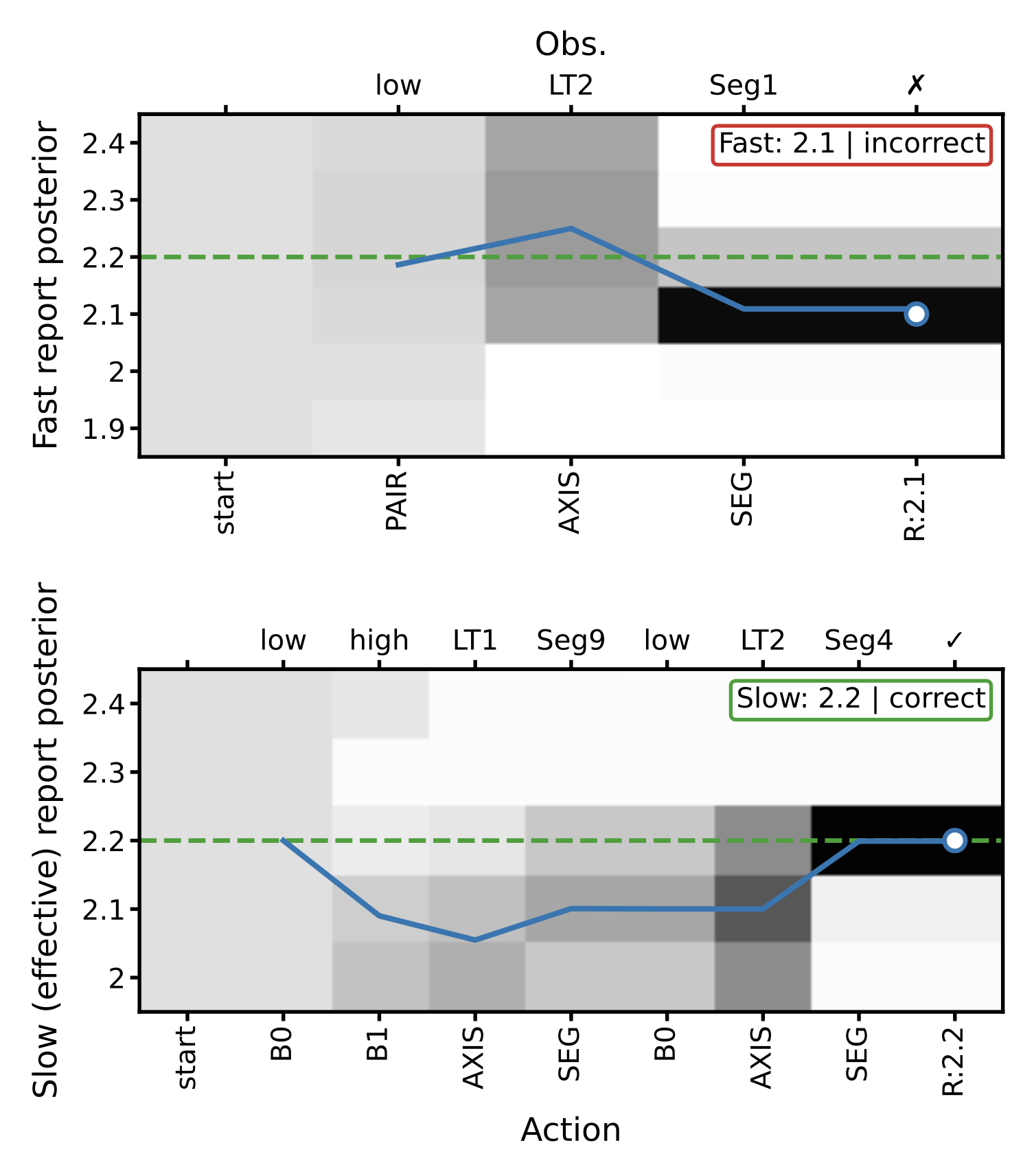}
    \caption{Example of heuristic failure in Fast model to early commit, while Slow model's deliberation leads to successful determination of bar average. In the top panel, the Fast model's posterior beliefs over the average bar value are plotted as a function of time show as the series of chosen actions. 
    Since the Slow model holds internal beliefs about each bar height and not the direct average, the bottom panel shows the believe average as a distribution marginalized over bar height beliefs at each timestep. 
    Both the model belief probabilities are plotted in greyscale, with higher probability darker. 
    The correct average (rounded to one decimal place) is plotted in green dashes and the maximum probability report belief at each timestep is plotted in blue. 
    }
    \label{fig:heuristic_failure_trace}
\end{figure}

Lets examine the illustrative example of Padilla \emph{et al.}, with \texttt{bar\_heights} $= (2.4, 1.9)$. With the above parameters and random \texttt{seed} $= 2$, the Fast agent mistakenly reports an average of $2.1$, while the answer rounded to one decimal place is $2.2$. Cognitive traces for both models are plotted in Fig.\ \ref{fig:heuristic_failure_trace}. The upper panel visualizes the Fast model's posterior belief over the average bar probability. The bottom panel shows the Slow model's effective belief over bar average, but this values is computed as a marginal distribution over the direct bar value beliefs kept by the Slow model. 

In this example, the Fast model starts by with the \texttt{LOOK\_PAIR} action which returns the observation \texttt{LowerHalf}, indicating the bar top falls in the lower half of the tick interval. Next the Fast agent selects action \texttt{LOOK\_TICK}, which reveals the lower bound of the interval is 2. At timestep 3 the agent selects action \texttt{LOOK\_SEGMENT} to reveal an observation of \texttt{Segment1} indicating approximate alignment with value 2.1. This incorrect result is due to observation noise on the imprecise segment observation. But the agent decides at timestep 4 it is certain enough to report an incorrect average bar value of 2.1.

The Slow agent is forced into a longer sequence of actions due to its structure. It first takes action \texttt{LOOK\_BAR\_0} and receives an observation of \texttt{LowerHalf}. It then selects action \texttt{LOOK\_BAR\_1} receiving observation \texttt{UpperHalf}, which strays from the optimal course of policies because the agent can only plan 4 timesteps in advance. At timestep 2, the agent proceeds with the intended policy sequence, selecting \texttt{LOOK\_TICK} and \texttt{LOOK\_SEGMENT} while anchored on bar 1. Finally the Slow agent returns to bar 0 with \texttt{LOOK\_BAR\_0}, eventually receiving correct observations from \texttt{LOOK\_SEGMENT} and reporting the correct average of 2.2. 

\begin{figure}
    \centering
    \includegraphics[width=\columnwidth]{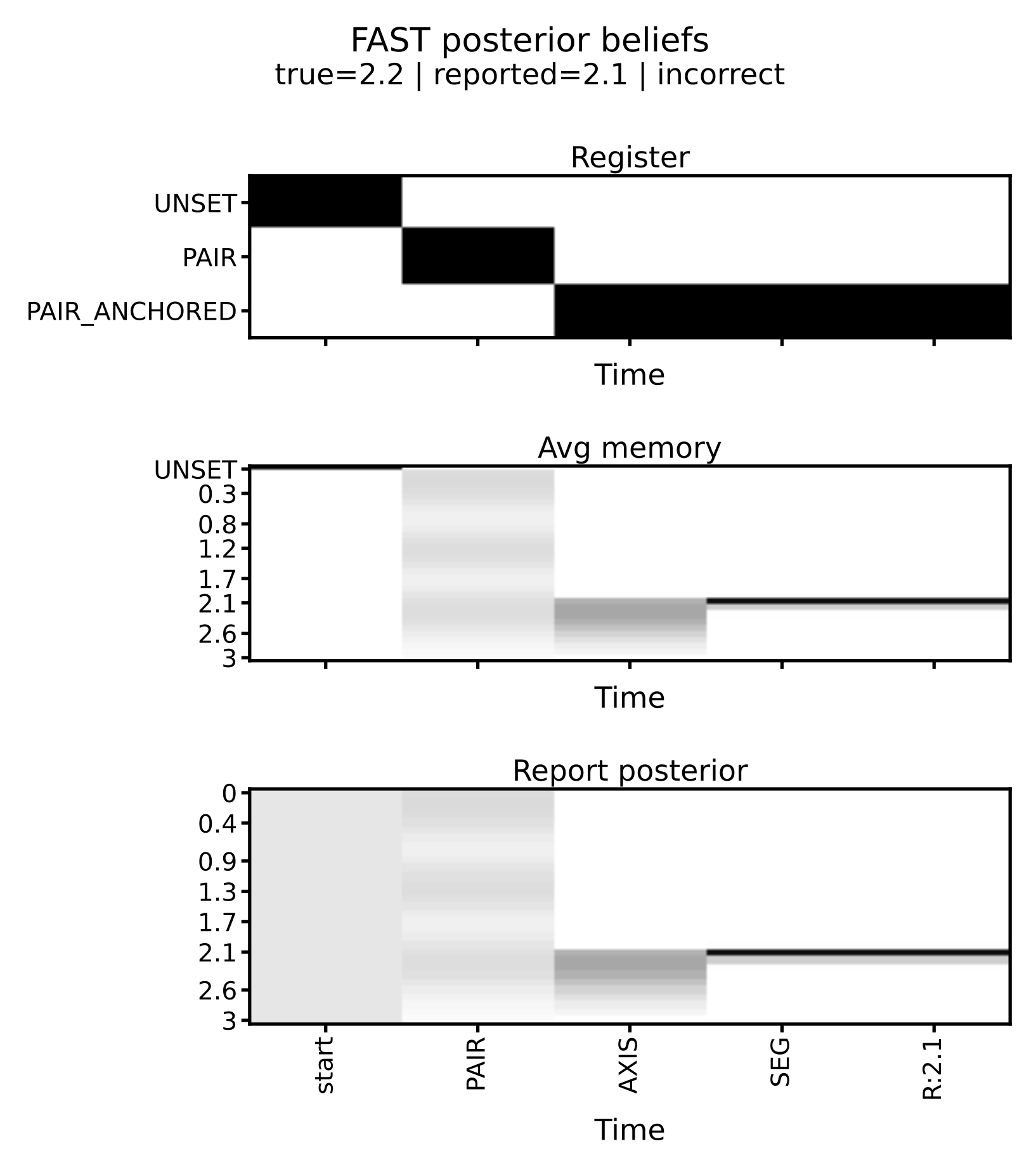}
    \caption{Fast model posterior beliefs from example test case of \texttt{bar\_heights} $= (2.4, 1.9)$, parameters from Table \ref{tab:appendix_defaults}, and random \texttt{seed} $= 2$.}
    \label{fig:fast_trace}
\end{figure}

\begin{figure}[t!]
    \centering
    \includegraphics[width=\columnwidth]{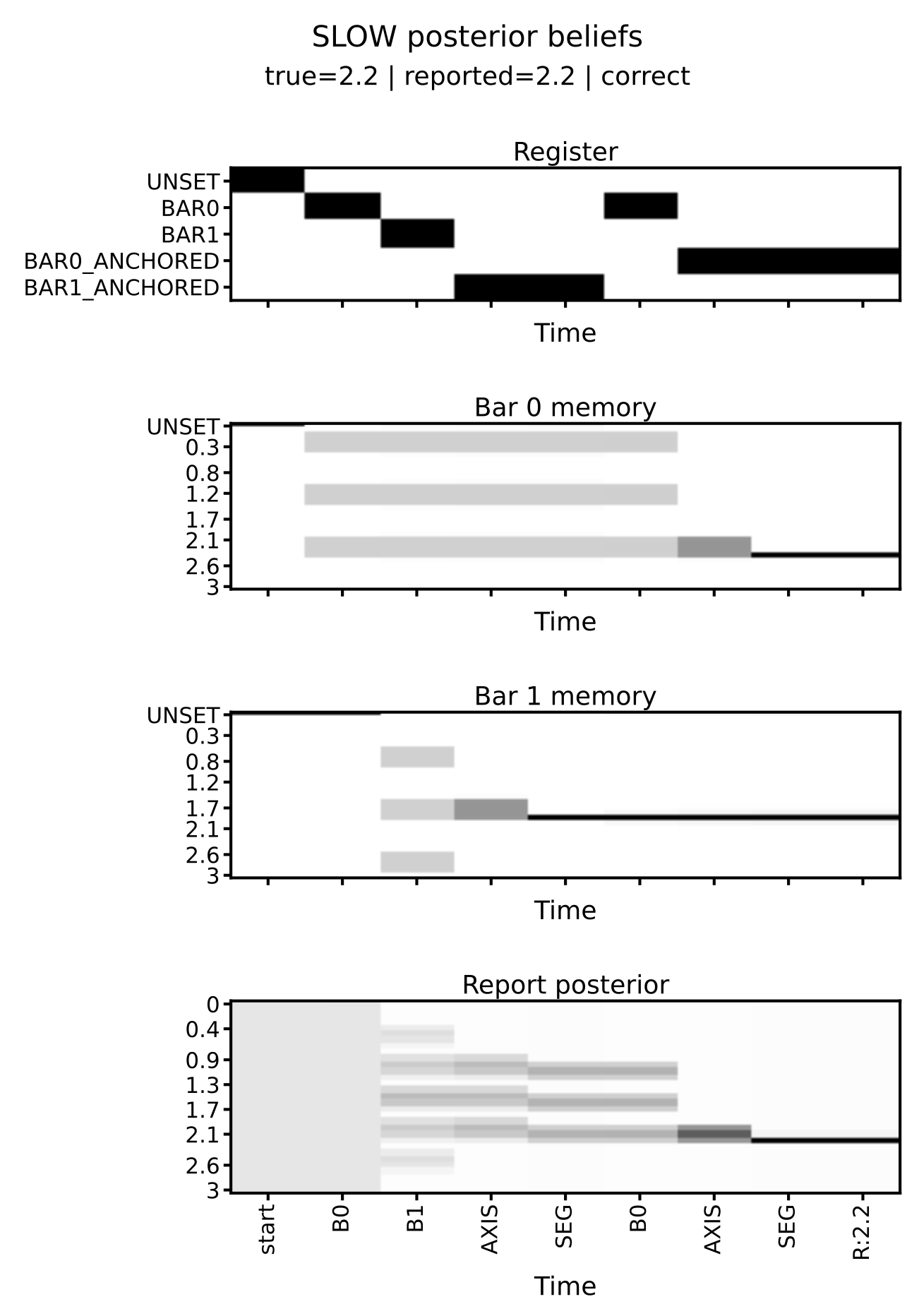}
    \caption{Slow model posterior beliefs from example test case of \texttt{bar\_heights} $= (2.4, 1.9)$, parameters from Table \ref{tab:appendix_defaults}, and random \texttt{seed} $= 2$.}
    \label{fig:slow_trace}
\end{figure}

For a comprehensive visualization of each agents posterior beliefs from this example case, see Fast model trace in Fig.~\ref{fig:fast_trace} and Slow model trace in Fig.~\ref{fig:slow_trace}.

\subsection{Memory and Tick-bias experiments: additional details}

\begin{figure}
    \centering
    \includegraphics[width=\columnwidth]{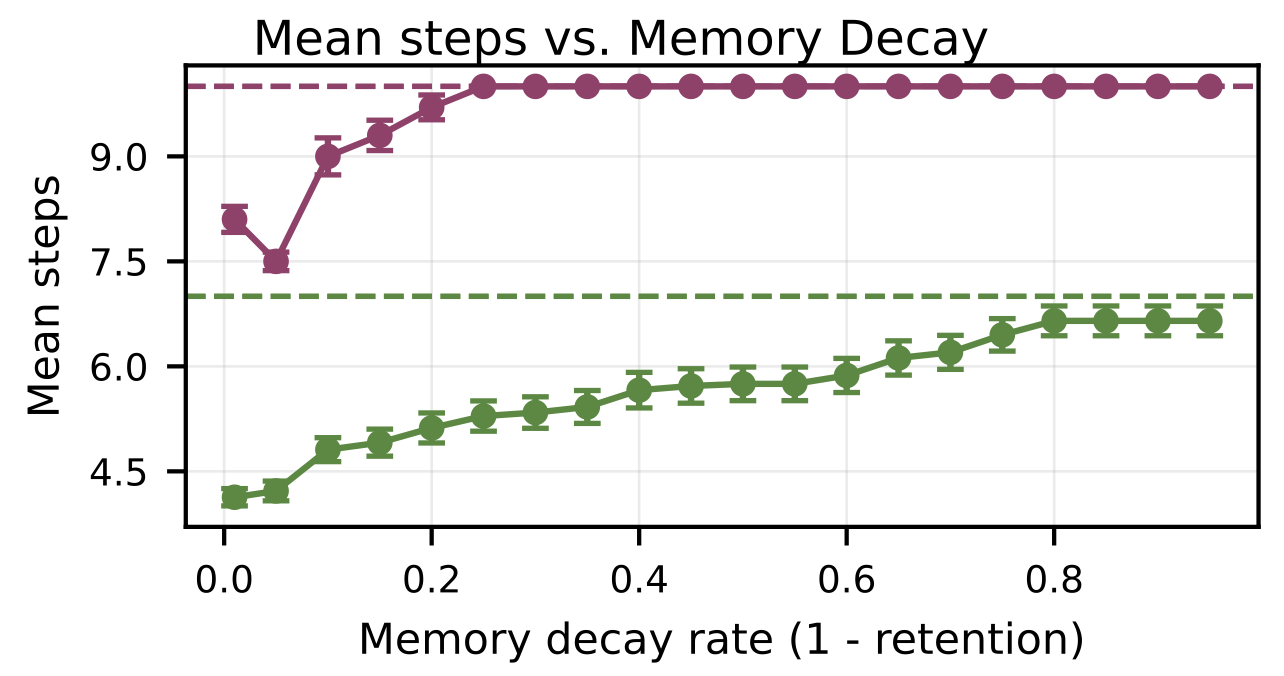}
    \caption{
    Average number of time steps used by both models over variation in memory decay rate. Slow model results are plotted in purple and Fast model results are plotted in green. Maximum times before forced decisions are plotted as dashed lines.
    }
    \label{fig:memory_steps}
\end{figure}

\begin{figure}
    \centering
    \includegraphics[width=\columnwidth]{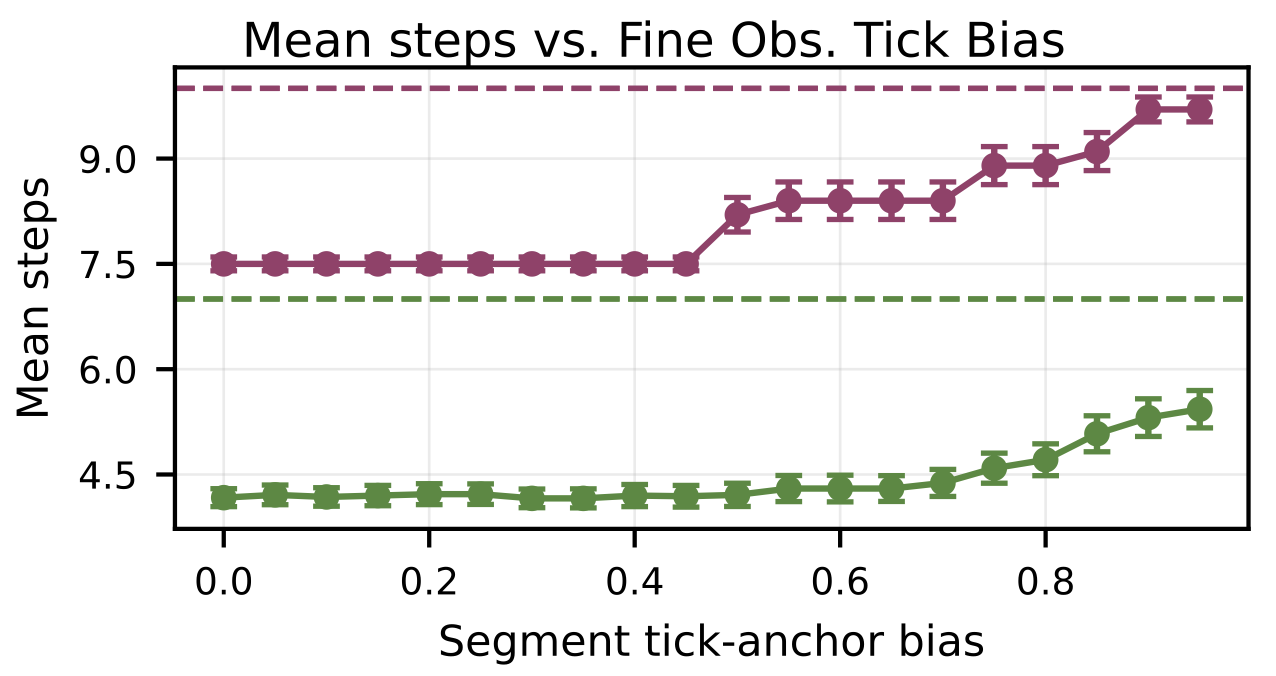}
    \caption{ 
    Average number of time steps used by both models over variation in tick bias applied to the segment refinement cue. Slow model results are plotted in purple and Fast model results are plotted in green. Maximum times before forced decisions are plotted as dashed lines.}
    \label{fig:tick_steps}
\end{figure}

For the memory and tick bias experiments, we gain intuition from examining the mean time steps taken by each agent. These are presented in Fig.~\ref{fig:memory_steps} and Fig.~\ref{fig:tick_steps}, respectively.

\section{Shared Formalism and Approximations}
\label{app:formalism}

\subsection{Canonical discrete-state formulation}
Here we elaborate on the mathematics of applied Active Inference. Let hidden states at time \(t\) be denoted \(s_t\), observations \(o_t\), and policies \(\pi=(u_t,u_{t+1},\dots,u_{t+H-1})\). In canonical discrete-state Active Inference, the generative model is specified by likelihood matrices \(A\), transition matrices \(B\), prior preferences \(C\), and initial-state priors \(D\), as defined in Eq.~\ref{eq:abcd}.
Writing \(\mathbf{s}_t\) for the posterior belief vector over hidden states and \(\mathbf{o}_t^{(m)}\) for a one-hot encoding of the realized outcome in modality \(m\), the predictive prior is
\begin{equation}
\tilde{\mathbf{s}}_t
=
\begin{cases}
\mathbf{D}, & t=1,\\[4pt]
\mathbf{B}^{(u_{t-1})}\mathbf{s}_{t-1}, & t>1.
\end{cases}
\label{eq:app_predictive_prior}
\end{equation}
The variational free energy at time \(t\) can then be written as
\begin{equation}
F_t(\mathbf{s}_t)
=
\mathbf{s}_t \cdot \ln \mathbf{s}_t
-
\mathbf{s}_t \cdot \ln \tilde{\mathbf{s}}_t
-
\sum_{m=1}^{M}
\mathbf{s}_t \cdot \ln\!\big(\mathbf{A}^{(m)\top}\mathbf{o}_t^{(m)}\big),
\label{eq:app_free_energy_matrix}
\end{equation}
where logarithms act elementwise. Minimizing \(F_t\) yields the standard discrete-state update
\begin{equation}
\mathbf{s}_t
=
\sigma\!\left(
\ln \tilde{\mathbf{s}}_t
+
\sum_{m=1}^{M}
\ln\!\big(\mathbf{A}^{(m)\top}\mathbf{o}_t^{(m)}\big)
\right),
\label{eq:app_state_update}
\end{equation}
with \(\sigma(\cdot)\) the softmax operator.

For a candidate policy \(\pi\), predicted future states and outcomes are generated recursively by
\begin{align}
\mathbf{s}_{\tau+1}^{\pi} &= \mathbf{B}^{(u_\tau)}\mathbf{s}_{\tau}^{\pi},
\qquad \mathbf{s}_t^\pi = \mathbf{s}_t, \\
\mathbf{o}_{\tau}^{(m),\pi} &= \mathbf{A}^{(m)}\mathbf{s}_{\tau}^{\pi}.
\end{align}

Policies are scored by expected free energy \(G(\pi)\).
Using preference vectors \(\mathbf{C}_\tau^{(m)}\) over outcomes, one convenient matrix form of expected free energy is
\begin{equation}
G(\pi)
=
\sum_{\tau=t}^{t+H-1}
\left[
\sum_{m=1}^{M}
\mathbf{o}_{\tau}^{(m),\pi}\!\cdot
\ln \big(
\mathbf{o}_{\tau}^{(m),\pi}
-
\mathbf{C}_{\tau}^{(m)}
\big)
+
\sum_{m=1}^{M}
\mathbf{s}_{\tau}^{\pi}\!\cdot\!\mathbf{H}^{(m)}
\right],
\label{eq:efe_matrix}
\end{equation}
where \(\mathbf{H}^{(m)}\) is the ambiguity vector of modality \(m\), with components
\begin{equation}
H^{(m)}_j
=
-\sum_i A^{(m)}_{ij}\ln A^{(m)}_{ij}.
\label{eq:ambiguity_vector}
\end{equation}
The first term in \eqref{eq:efe_matrix} penalizes policies expected to produce non-preferred outcomes, while the second penalizes policies expected to sample ambiguous observations.
Policies are then selected according to
\begin{equation}
q(\pi)
=
\sigma\!\big(
\ln p(\pi) - \gamma G(\pi)
\big),
\label{eq:policy_posterior_matrix}
\end{equation}
with the next action taken as the first element of the selected policy.
In this form, action selection is explicit in the \(A/B/C/D\) matrices: \(B\) predicts future hidden states under candidate actions, \(A\) maps those predicted states to outcomes, \(C\) scores those outcomes according to task preferences, and \(D\) initializes inference at the start of a trial.

\subsection{Known-action operator form}
The final Fast and Slow models use an equivalent \emph{known-action} reduction. Rather than representing the currently selected perceptual operation as an additional hidden-state factor, the model conditions directly on the chosen action \(u_t\). The one-step generative model is therefore
\begin{equation}
p(o_t,s_{t+1}\mid s_t,u_t)
=
p(o_t\mid s_t,u_t)\,p(s_{t+1}\mid s_t,u_t).
\label{eq:app_known_action_model}
\end{equation}
For each action \(u\in\mathcal U\), we define an action-conditioned transition operator \(B^{(u)}\) and action-conditioned likelihood operators \(A^{(m,u)}\), with
\begin{equation}
A^{(m,u)}_{ij}
=
p(o_t^{(m)}=i \mid s_t=j,u_t=u).
\end{equation}
Writing \(q_{t-1}\) for the posterior over hidden states at time \(t-1\), the predictive update under action \(u_t\) is $B^{(u_t)} q_{t-1}$. For realized observation \(o_t=(o_t^{(1)},\dots,o_t^{(M)})\), define the diagonal likelihood operator
\begin{equation}
\Lambda^{(u_t)}(o_t)
=
\prod_{m=1}^{M}
\mathrm{Diag}\!\left(A^{(m,u_t)}_{o_t^{(m)},:}\right).
\label{eq:app_likelihood_operator}
\end{equation}
The posterior update is then
\begin{equation}
q_t
=
\frac{
\Lambda^{(u_t)}(o_t)\,B^{(u_t)} q_{t-1}
}{
\mathbf{1}^\top \Lambda^{(u_t)}(o_t)\,B^{(u_t)} q_{t-1}
},
\label{eq:app_posterior_update}
\end{equation}
or equivalently
\begin{equation}
q_t
=
\sigma\!\left(
\ln \Lambda^{(u_t)}(o_t)
+
\ln\!\big(B^{(u_t)}q_{t-1}\big)
\right).
\label{eq:app_posterior_update_softmax}
\end{equation}
This controlled Bayes-filter form is the state update shared by both Fast and Slow models.

\subsection{Approximate policy scoring}
\label{app:approx_policy_scoring}

The state update in Eq.~\eqref{eq:app_posterior_update} is exact with
respect to the chosen action-conditioned symbolic generative model. The
approximations enter through the finite hidden-state representation and the
tractable evaluation of short candidate policies.

Canonical Active Inference selects policies with low expected free energy
\(G(\pi)\). In the implementation, we assign each candidate policy a
higher-is-better score
\begin{equation}
S(\pi)
=
-\widehat{G}(\pi)
=
U(\pi)+I(\pi)-\kappa(\pi),
\label{eq:app_neg_efe_score}
\end{equation}
where \(U(\pi)\) is expected preference value, \(I(\pi)\) is expected
information gain, and \(\kappa(\pi)\) is explicit effort cost. Thus larger
scores correspond to lower approximate expected free energy.

Let \(q_t\) denote the current categorical belief vector over hidden states.
For candidate action \(a\), the predicted belief before observing the next
outcome is
\begin{equation}
q_t^a
=
B_a q_t ,
\label{eq:app_action_predicted_belief}
\end{equation}
where \(B_a\) is the action-conditioned transition matrix. The predicted
observation vector is
\begin{equation}
o_t^a
=
A_a q_t^a ,
\label{eq:app_pred_obs}
\end{equation}
where \(A_a\) is the action-conditioned likelihood matrix. When the model has
multiple observation modalities, Eq.~\eqref{eq:app_pred_obs} is evaluated
separately for each modality, using the corresponding likelihood matrix
\(A_a^{(m)}\).

For a possible observation \(o=i\), the hypothetical posterior used during
policy evaluation is the normalized elementwise product
\begin{equation}
q_t^{a,i}
=
\mathrm{norm}
\!\left(
A_a[i,:] \odot q_t^a
\right),
\label{eq:app_hypothetical_posterior}
\end{equation}
where \(A_a[i,:]\) is the likelihood row for observation \(i\), \(\odot\)
denotes elementwise multiplication, and \(\mathrm{norm}(v)=v/(\mathbf{1}^\top v)\).
This is the same controlled Bayes update as
Eq.~\eqref{eq:app_posterior_update}, but used prospectively during policy
evaluation.

The one-step preference term is
\begin{equation}
U_t(a)
=
C_a^\top o_t^a ,
\label{eq:app_utility_score}
\end{equation}
where \(C_a\) is the preference vector over predicted observations. In the
present simulations, preferences are applied only to the feedback modality
\(\{\texttt{Null},\texttt{Incorrect},\texttt{Correct}\}\), with
\begin{equation}
C_{\mathrm{fb}}
=
(0,\lambda_-,\lambda_+)
=
(0,-80,12).
\label{eq:app_pref_vector}
\end{equation}
All non-feedback modalities have zero preference scores.

The epistemic term is computed over the report distribution. Let \(P\) denote
the deterministic projection from hidden states to the one-decimal report grid.
For any hidden-state belief \(q\), the corresponding report belief is
\begin{equation}
q_r = Pq .
\label{eq:app_report_projection}
\end{equation}
The expected information gain from action \(a\) is then
\begin{equation}
I_t(a)
=
H\!\left[Pq_t^a\right]
-
\sum_i
o_t^a(i)\,
H\!\left[Pq_t^{a,i}\right].
\label{eq:app_information_score}
\end{equation}
Thus the agent favors actions expected to reduce uncertainty about the value it
will ultimately report. The projection \(P\) differs between the two models:
for the Fast model it maps the average-memory state directly to the report
grid, whereas for the Slow model it maps pairs of remembered bar heights to
their rounded average.

The one-step effort term is
\begin{equation}
\kappa_t(a)
=
\begin{cases}
0, & a=\texttt{REPORT}_r,\\
\kappa_{\mathrm{nr}}, & \text{otherwise}.
\end{cases}
\label{eq:app_action_cost}
\end{equation}
The immediate action score is therefore
\begin{equation}
S_t(a)
=
U_t(a)+I_t(a)-\kappa_t(a).
\label{eq:app_action_score}
\end{equation}

For a candidate policy \(\pi=(a_1,\ldots,a_K)\), scores are accumulated until
the first report action:
\begin{align}
U(\pi)
&=
\sum_{k=1}^{K_\pi} U_{t+k-1}(a_k),
\\
I(\pi)
&=
\sum_{k=1}^{K_\pi} I_{t+k-1}(a_k),
\\
\kappa(\pi)
&=
\sum_{k=1}^{K_\pi} \kappa_{t+k-1}(a_k),
\end{align}
so that
\begin{equation}
S(\pi)
=
\sum_{k=1}^{K_\pi} S_{t+k-1}(a_k)
=
U(\pi)+I(\pi)-\kappa(\pi).
\label{eq:app_policy_score}
\end{equation}
Here \(K_\pi\) is the effective policy length after truncating at the first
report.

The posterior over policies is then
\begin{equation}
q(\pi)
\propto
p(\pi)\,\exp\!\big(\gamma S(\pi)\big),
\label{eq:app_policy_softmax}
\end{equation}
where \(\gamma\) is the policy precision. With a uniform prior over the finite
policy library, this is equivalent to applying a softmax to the policy scores.
Because \(S(\pi)=-\widehat{G}(\pi)\), this remains consistent with the
canonical relation \(q(\pi)\propto \exp[-\gamma G(\pi)]\).

The reported experiments use rollout evaluation: after scoring an action, the
next simulated belief is the predictive belief \(q_t^a\), avoiding explicit
branching over all possible future observations. The implementation also
supports branching evaluation, in which each possible observation \(i\) is
weighted by \(o_t^a(i)\), updated using
Eq.~\eqref{eq:app_hypothetical_posterior}, and recursively evaluated for the
remaining actions. Branching is closer to finite-horizon expected free energy,
whereas rollout is computationally cheaper.


\subsection{Additional approximations}
\paragraph{Restricted policy library and short horizon.}
Rather than evaluating all possible action sequences, both models score a hand-designed library of plausible 4 step policies. This is both a tractability approximation and a psychologically motivated assumption that human observers bring a small repertoire of useful routines to familiar chart-reading tasks.

\paragraph{Report-targeted epistemics.}
Epistemic value may be computed either over the full hidden state or over the induced one-decimal report distribution. The Fast and Slow models use report-targeted epistemics by default, because the behaviorally relevant question is which one-decimal answer the model will eventually give.

\paragraph{Rollout versus branching.}
The implemented models support both \emph{rollout} and \emph{branching} evaluation of future actions. Branching explicitly averages over future observations and is the closer analogue of finite-horizon expected free energy. Rollout propagates only predictive beliefs and is cheaper computationally.

\paragraph{Deterministic action selection and instantaneous report.}
Policy beliefs are computed as a softmax over policy scores, but both models usually execute the first action of the highest-scoring policy. Report actions may also be treated as instantaneous commitments that do not incur an additional memory-decay step before feedback.

Taken together, these choices yield compact, task-adapted, Active-Inference-inspired process models. The main approximations concern planning and state-space reduction; the central cognitive hypotheses of the paper---bar-wise versus compressed representation, working-memory decay, and tick-salience bias---are not conveniences around the formalism, but the substantive mechanisms under test.

\section{Shared Task and Cue Construction}
\label{app:shared_task}

\subsection{Task, grids, and report criterion}
We consider two-bar charts with latent bar heights
\begin{equation}
h_0,h_1 \in \mathcal{H},
\qquad
\mathcal{H}
=
\{0,\Delta_h,2\Delta_h,\dots,n_{\mathrm{ticks}}-1\},
\label{eq:app_height_grid}
\end{equation}
where \(\Delta_h\) is the internal height discretization and must land exactly on the top tick.

The behaviorally relevant report space is the one-decimal grid
\begin{equation}
\mathcal{R}
=
\{0,0.1,0.2,\dots,n_{\mathrm{ticks}}-1\}.
\label{eq:app_report_grid}
\end{equation}
The correct response for a stimulus is the rounded target
\begin{equation}
r^\star
=
\mathrm{round}_{0.1}\!\left(\frac{h_0+h_1}{2}\right),
\label{eq:app_true_report}
\end{equation}
where \(\mathrm{round}_{0.1}\) denotes half-up rounding to the nearest tenth.

The Slow model stores bar-specific memories on \(\mathcal{H}\) and projects them onto \(\mathcal{R}\) only at report time. The Fast model stores its average memory directly on \(\mathcal{R}\).

\subsection{Shared action and observation vocabulary}
Both models use the same staged perceptual logic. A first action samples a coarse cue, a second anchors that cue to the axis, a third refines it within the relevant interval, and a final report action commits to a one-decimal answer.

The shared observation tuple is
\begin{equation}
o_t
=
\bigl(
o_t^{\mathrm{rel}},
o_t^{\mathrm{axis}},
o_t^{\mathrm{seg}},
o_t^{\mathrm{fb}}
\bigr).
\label{eq:app_obs_tuple}
\end{equation}
The four modalities are
\begin{align}
o_t^{\mathrm{rel}}
&\in
\{\texttt{Null},\texttt{OnTick},\texttt{LowerHalf},\texttt{Midpoint},\texttt{UpperHalf}\},\\
o_t^{\mathrm{axis}}
&\in
\{\texttt{Null}\}
\cup
\{\texttt{Tick}_k\}_{k=0}^{n_{\mathrm{ticks}}-1}
\cup
\{\texttt{LowerTick}_k\}_{k=0}^{n_{\mathrm{ticks}}-2},\\
o_t^{\mathrm{seg}}
&\in
\{\texttt{Null}\}\cup\{\texttt{Seg}_j\}_{j=0}^{9},\\
o_t^{\mathrm{fb}}
&\in
\{\texttt{Null},\texttt{Incorrect},\texttt{Correct}\}.
\end{align}

For any value \(v\), define the coarse relative-position classifier
\begin{equation}
\rho(v)
=
\begin{cases}
\texttt{OnTick}, & \mathrm{frac}(v)=0,\\
\texttt{LowerHalf}, & 0<\mathrm{frac}(v)<0.5,\\
\texttt{Midpoint}, & \mathrm{frac}(v)=0.5,\\
\texttt{UpperHalf}, & 0.5<\mathrm{frac}(v)<1,
\end{cases}
\label{eq:app_rel_classifier}
\end{equation}
the local axis-anchor classifier
\begin{equation}
\alpha(v)
=
\begin{cases}
\texttt{Tick}_k, & v=k,\\
\texttt{LowerTick}_k, & k<v<k+1,\quad k=\lfloor v\rfloor,
\end{cases}
\label{eq:app_axis_classifier}
\end{equation}
and the within-interval segment index
\begin{equation}
\eta(v)\in\{0,\dots,9\}.
\label{eq:app_seg_classifier}
\end{equation}

The combined cue triple \((\rho,\alpha,\eta)\) must uniquely identify each allowed latent value on the chosen grid. In the implemented models, this identifiability condition is explicitly checked at initialization for the bar-height grid in the Slow model and for the report grid in the Fast model.

\subsection{Exact cue likelihoods}
For either model, let \(\mathcal{G}\) denote the relevant value grid: \(\mathcal{G}=\mathcal{H}\) for bar-specific Slow memories and \(\mathcal{H}=\mathcal{R}\) for the Fast average memory. Let \(\mathcal{H}=\{g_\ell\}\) in increasing order.

We define a normalized Gaussian on the discrete grid by
\begin{equation}
\phi_\sigma(g_\ell;\mu)
=
\frac{
\exp\!\left[-\frac12\left(\frac{g_\ell-\mu}{\sigma}\right)^2\right]
}{
\sum_{\ell'}\exp\!\left[-\frac12\left(\frac{g_{\ell'}-\mu}{\sigma}\right)^2\right]
},
\label{eq:app_grid_gaussian}
\end{equation}
with the limiting case \(\sigma\to 0\) taken as a deterministic snap to the nearest grid value.

For a value \(v\) and nearest integer tick
\begin{equation}
\tau(v)=\mathrm{clip}\!\bigl(\lfloor v+0.5\rfloor,\,0,\,n_{\mathrm{ticks}}-1\bigr),
\label{eq:app_nearest_tick}
\end{equation}
a tick-anchored perceptual shift is written
\begin{equation}
\tilde v
=
(1-\lambda_{\mathrm{tick}})v+\lambda_{\mathrm{tick}}\tau(v).
\label{eq:app_tick_shift}
\end{equation}

The coarse cue likelihood is then obtained by aggregating grid mass into the five relative-position classes:
\begin{equation}
p_{\mathrm{rel}}(o^{\mathrm{rel}} \mid v;\sigma_{\mathrm{coarse}},\lambda_{\mathrm{tick}})
=
\sum_{\ell:\rho(g_\ell)=o^{\mathrm{rel}}}
\phi_{\sigma_{\mathrm{coarse}}}(g_\ell;\tilde v).
\label{eq:app_rel_likelihood}
\end{equation}
In both generative models, \(\lambda_{\mathrm{tick}}=0\) within the likelihood while environment-side cue is biased. Thus the agents do not know they succumb to tick-salience bias. 

The axis cue is deterministic:
\begin{equation}
p_{\mathrm{axis}}(o^{\mathrm{axis}}\mid v)
=
\delta\!\bigl(o^{\mathrm{axis}},\alpha(v)\bigr).
\label{eq:app_axis_likelihood}
\end{equation}

The refined segment cue is obtained analogously by aggregating grid mass into the 10 within-interval segment bins:
\begin{equation}
p_{\mathrm{seg}}(o^{\mathrm{seg}} \mid v;\sigma_{\mathrm{seg}},\lambda_{\mathrm{tick}})
=
\sum_{\ell:\eta(g_\ell)=o^{\mathrm{seg}}}
\phi_{\sigma_{\mathrm{seg}}}(g_\ell;\tilde v),
\label{eq:app_seg_likelihood}
\end{equation}
again with \(\lambda_{\mathrm{tick}}=0\) in the current planner-side comparison models. Thus, the current Fast/Slow comparison files instantiate process--model mismatch by allowing the environment to bias the cue centroid while the agent's own cue tables remain unbiased.

For the reported experiments, the coarse cue width is \(\sigma_{\mathrm{coarse}}=\texttt{bar\_obs\_sigma}\) in the Slow model and \(\sigma_{\mathrm{coarse}}=\texttt{pair\_obs\_sigma}\) in the Fast model. The segment cue width is \(\sigma_{\mathrm{seg}}=\texttt{segment\_sigma}\) in both.

\section{Fast and Slow Architectures}
\label{app:architectures}

Both models are known-action Active Inference agents for the same bar-chart
averaging task. They share the same overall control logic: the agent gathers a
coarse cue, anchors that cue to the axis, optionally refines it with a
within-interval segment cue, and then commits to a report. They also share the
same general treatment of working memory: populated memory states are retained
or diffused by a decay operator between actions, \texttt{UNSET} represents the
absence of task-relevant memory, and report actions may be treated as
instantaneous commitments when \texttt{report\_action\_instant} is enabled.
The principal architectural difference is that the Slow agent maintains two
bar-specific memories, whereas the Fast agent maintains one directly encoded
average memory.

\subsection{Slow model}
\label{app:slow_full}

\subsubsection{Hidden state and action set}
The Slow agent hidden state is
\begin{equation}
x_t^{\mathrm{slow}}
=
\bigl(
r_t,\,
m_t^{(0)},\,
m_t^{(1)}
\bigr),
\label{eq:app_slow_state}
\end{equation}
where
\begin{align}
r_t
&\in
\{\texttt{UNSET},\texttt{BAR0},\texttt{BAR1},\texttt{BAR0\_ANCHORED},\texttt{BAR1\_ANCHORED}\},\\
m_t^{(0)} &\in \{\texttt{UNSET}\}\cup\mathcal{H},\\
m_t^{(1)} &\in \{\texttt{UNSET}\}\cup\mathcal{H}.
\end{align}
The register indicates which bar is currently selected and whether that bar has
been anchored to the axis. The two memory factors store bar-specific
working-memory traces on the allowed height grid \(\mathcal H\).

The Slow action set is
\begin{multline}
\mathcal{U}_{\mathrm{slow}}
=
\{
\texttt{LOOK\_BAR0},
\texttt{LOOK\_BAR1},
\texttt{LOOK\_AXIS},
\texttt{LOOK\_SEGMENT}
\}
\\\cup
\{
\texttt{REPORT}_r : r\in\mathcal{R}
\},
\label{eq:app_slow_actions}
\end{multline}
with intended bar-wise subroutine
\begin{equation}
\texttt{LOOK\_BAR}_i
\rightarrow
\texttt{LOOK\_AXIS}
\rightarrow
\texttt{LOOK\_SEGMENT},
\label{eq:app_slow_subroutine}
\end{equation}
applied first to one bar and then to the other, followed by a report action.

\subsubsection{Initial state and transitions}
The Slow agent is initialized at
\begin{multline}
q_0^{\mathrm{slow}}(r,m^{(0)},m^{(1)})
\\=
\delta(r=\texttt{UNSET})\,
\delta(m^{(0)}=\texttt{UNSET})\,
\delta(m^{(1)}=\texttt{UNSET}).
\end{multline}

Let \(W\) denote the bar-memory write-or-preserve operator and \(D\) the
memory-decay operator. Then
\begin{equation}
(r,m^{(0)},m^{(1)})
\mapsto
(\texttt{BAR0},\,Wm^{(0)},\,Dm^{(1)})
\label{eq:app_slow_bar0_transition}
\end{equation}
under \(\texttt{LOOK\_BAR0}\), and
\begin{equation}
(r,m^{(0)},m^{(1)})
\mapsto
(\texttt{BAR1},\,Dm^{(0)},\,Wm^{(1)})
\label{eq:app_slow_bar1_transition}
\end{equation}
under \(\texttt{LOOK\_BAR1}\).

Under \(\texttt{LOOK\_AXIS}\), both memory factors decay and the register is
anchored:
\begin{equation}
\texttt{BAR0}\mapsto\texttt{BAR0\_ANCHORED},
\qquad
\texttt{BAR1}\mapsto\texttt{BAR1\_ANCHORED},
\label{eq:app_slow_axis_transition}
\end{equation}
with anchored states preserved. Under \(\texttt{LOOK\_SEGMENT}\), the register
is preserved and both memories decay. If
\texttt{report\_action\_instant} is enabled, report actions do not incur an
additional decay step before feedback.

\subsubsection{Observation model}
The Slow agent uses action-conditioned likelihood tables
\begin{equation}
O^{\mathrm{slow}}_a(o\mid r,m^{(0)},m^{(1)}).
\end{equation}
For \(\texttt{LOOK\_BAR0}\), only the first bar memory is informative:
\begin{equation}
p(o^{\mathrm{rel}}\mid x,\texttt{LOOK\_BAR0})
=
\begin{cases}
\begin{aligned}
&\delta(o^{\mathrm{rel}},\texttt{Null}), 
\\&\quad m^{(0)}=\texttt{UNSET},
\end{aligned}
\\
\begin{aligned}
&p_{\mathrm{rel}}(o^{\mathrm{rel}}\mid h(m^{(0)});\texttt{bar\_obs\_sigma},0), 
\\&\quad m^{(0)}\neq\texttt{UNSET},
\end{aligned}
\end{cases}
\end{equation}
and analogously for \(\texttt{LOOK\_BAR1}\) using \(m^{(1)}\). In the current
implementation, the agent-side coarse cue omits explicit tick bias even when
the environment-side cue is biased.

For \(\texttt{LOOK\_AXIS}\), the axis cue is deterministic whenever the
register indicates a selected or anchored bar and the corresponding memory is
set:
\begin{equation}
p(o^{\mathrm{axis}}\mid x,\texttt{LOOK\_AXIS})
=
\delta\!\bigl(o^{\mathrm{axis}},\alpha(h(m^{(i)}))\bigr),
\end{equation}
with \texttt{Null} otherwise.

For \(\texttt{LOOK\_SEGMENT}\), the segment cue is informative only in the
anchored state associated with the currently selected bar:
\begin{multline}
p(o^{\mathrm{seg}}\mid x,\texttt{LOOK\_SEGMENT})
\\=
\begin{cases}
\begin{aligned}
&p_{\mathrm{seg}}(o^{\mathrm{seg}}\mid h(m^{(0)});\texttt{segment\_sigma},0), 
\\&\quad r=\texttt{BAR0\_ANCHORED},
\end{aligned}\\
\begin{aligned}
&p_{\mathrm{seg}}(o^{\mathrm{seg}}\mid h(m^{(1)});\texttt{segment\_sigma},0), 
\\&\quad r=\texttt{BAR1\_ANCHORED},
\end{aligned}\\
\begin{aligned}
&\delta(o^{\mathrm{seg}},\texttt{Null}), 
\\&\quad \text{otherwise}.
\end{aligned}
\end{cases}
\end{multline}
Again, in the current implementation the agent-side segment cue omits explicit
tick bias even when the environment-side segment cue is biased, so the agent
does not represent itself as biased.

\subsubsection{Report projection and report feedback}
The Slow agent reports from its internal memory contents rather than directly
from the true bar heights. When both bar memories are populated, the implied
internal average is
\begin{equation}
\hat{a}(m^{(0)},m^{(1)})
=
\frac{h(m^{(0)})+h(m^{(1)})}{2},
\end{equation}
with implied one-decimal report
\begin{equation}
\hat{r}(m^{(0)},m^{(1)})
=
\mathrm{round}_{0.1}\!\bigl(\hat{a}(m^{(0)},m^{(1)})\bigr).
\end{equation}
The report projection
\begin{equation}
q_t^{\mathrm{rep}}(r)
=
P_{\mathrm{slow}}\,q_t^{\mathrm{slow}}
\end{equation}
maps each populated joint memory state to its corresponding one-decimal report
value and maps any state with at least one \texttt{UNSET} memory uniformly over
\(\mathcal R\).

For report feedback, \(\texttt{REPORT}_r\) yields \texttt{Incorrect} whenever
either bar memory is \texttt{UNSET}; otherwise it yields \texttt{Correct} iff
\(r=\hat{r}(m^{(0)},m^{(1)})\).

\subsection{Fast model}
\label{app:fast_full}

\subsubsection{Hidden state and action set}
The Fast agent hidden state is
\begin{equation}
x_t^{\mathrm{fast}}
=
\bigl(
r_t,\,
m_t^{(\mathrm{avg})}
\bigr),
\label{eq:app_fast_state}
\end{equation}
where
\begin{align}
r_t &\in \{\texttt{UNSET},\texttt{PAIR},\texttt{PAIR\_ANCHORED}\},\\
m_t^{(\mathrm{avg})} &\in \{\texttt{UNSET}\}\cup\mathcal{R}.
\end{align}
Thus, the Fast agent stores a single directly encoded average memory on the
one-decimal report grid.

Its action set is
\begin{multline}
\mathcal{U}_{\mathrm{fast}}
=
\{
\texttt{LOOK\_PAIR},
\texttt{LOOK\_AXIS},
\texttt{LOOK\_SEGMENT}
\}
\\\cup
\{
\texttt{REPORT}_r : r\in\mathcal{R}
\},
\label{eq:app_fast_actions}
\end{multline}
with intended routine
\begin{equation}
\texttt{LOOK\_PAIR}
\rightarrow
\texttt{LOOK\_AXIS}
\rightarrow
\texttt{LOOK\_SEGMENT}
\rightarrow
\texttt{REPORT}_r.
\label{eq:app_fast_subroutine}
\end{equation}

\subsubsection{Initial state and transitions}
The Fast agent starts from
\begin{equation}
q_0^{\mathrm{fast}}(r,m)
=
\delta(r=\texttt{UNSET})\,\delta(m=\texttt{UNSET}).
\end{equation}

Let \(W\) denote the average-memory write-or-preserve operator and \(D\) the
same general decay operator used above. Under \(\texttt{LOOK\_PAIR}\),
\begin{equation}
(r,m)\mapsto(\texttt{PAIR},Wm).
\end{equation}
Under \(\texttt{LOOK\_AXIS}\), memory decays and the register transitions from
\texttt{PAIR} to \texttt{PAIR\_ANCHORED}, with anchored states preserved. Under
\(\texttt{LOOK\_SEGMENT}\), the register is preserved and memory decays. As in
the Slow model, report actions may be treated as instantaneous commitments.

\subsubsection{Observation model}
The Fast agent uses action-conditioned likelihood tables
\begin{equation}
O^{\mathrm{fast}}_a(o\mid r,m^{(\mathrm{avg})}).
\end{equation}
For \(\texttt{LOOK\_PAIR}\), the relative cue depends only on the average
memory:
\begin{equation}
p(o^{\mathrm{rel}}\mid x,\texttt{LOOK\_PAIR})
=
\begin{cases}
\begin{aligned}
&\delta(o^{\mathrm{rel}},\texttt{Null}), 
\\&\quad m^{(\mathrm{avg})}=\texttt{UNSET},
\end{aligned}\\
\begin{aligned}
&p_{\mathrm{rel}}(o^{\mathrm{rel}}\mid m^{(\mathrm{avg})};\texttt{pair\_obs\_sigma},0),
\\&\quad m^{(\mathrm{avg})}\neq\texttt{UNSET}.
\end{aligned}
\end{cases}
\end{equation}
As in the Slow model, the agent-side coarse cue omits explicit tick bias even
when the environment-side cue is biased.

For \(\texttt{LOOK\_AXIS}\), the axis cue is deterministic whenever the
register is \texttt{PAIR} or \texttt{PAIR\_ANCHORED} and the average memory is
set:
\begin{equation}
p(o^{\mathrm{axis}}\mid x,\texttt{LOOK\_AXIS})
=
\delta\!\bigl(o^{\mathrm{axis}},\alpha(m^{(\mathrm{avg})})\bigr),
\end{equation}
with \texttt{Null} otherwise.

For \(\texttt{LOOK\_SEGMENT}\), the segment cue is informative only in the
anchored state:
\begin{multline}
p(o^{\mathrm{seg}}\mid x,\texttt{LOOK\_SEGMENT})
\\=
\begin{cases}
\begin{aligned}
&p_{\mathrm{seg}}(o^{\mathrm{seg}}\mid m^{(\mathrm{avg})};\texttt{segment\_sigma},0), 
\\&\quad r=\texttt{PAIR\_ANCHORED},
\end{aligned}
\\
\begin{aligned}
&\delta(o^{\mathrm{seg}},\texttt{Null}), 
\\&\quad \text{otherwise}.
\end{aligned}
\end{cases}
\end{multline}
Again, the agent-side segment cue omits explicit tick bias even when the
environment-side segment cue is biased.

\subsubsection{Report projection and report feedback}
Because the Fast average memory already lives on \(\mathcal R\), report
projection is simpler:
\begin{equation}
q_t^{\mathrm{rep}}(r)
=
P_{\mathrm{fast}}\,q_t^{\mathrm{fast}},
\end{equation}
where populated memory states map directly to their corresponding report values
and the \texttt{UNSET} state projects uniformly over \(\mathcal R\).

For report feedback, \(\texttt{REPORT}_r\) yields \texttt{Incorrect} whenever
the average memory is \(\texttt{UNSET}\); otherwise it yields \(\texttt{Correct}\)
iff \(r=m_t^{(\mathrm{avg})}\).

\subsection{Policy-template families}
\label{app:policy_templates}

Both agents use horizon-4 hand-designed policy libraries comprising
exploratory/refinement templates and report-ending templates. The report-ending
families are structurally similar across the two models: each includes a
full acquire-anchor-refine-report sequence, shorter refine-to-report
sequences, and a degenerate commit family consisting entirely of repeated
\(\texttt{REPORT}_r\) actions.

For the Slow agent, the exploratory templates alternate between the two
bar-specific subroutines:
\begin{align}
&[\texttt{LOOK\_BAR0},\texttt{LOOK\_AXIS},\texttt{LOOK\_SEGMENT},\texttt{LOOK\_BAR1}],\\
&[\texttt{LOOK\_BAR1},\texttt{LOOK\_AXIS},\texttt{LOOK\_SEGMENT},\texttt{LOOK\_BAR0}],\\
&[\texttt{LOOK\_AXIS},\texttt{LOOK\_SEGMENT},\texttt{LOOK\_BAR0},\texttt{LOOK\_AXIS}],\\
&[\texttt{LOOK\_AXIS},\texttt{LOOK\_SEGMENT},\texttt{LOOK\_BAR1},\texttt{LOOK\_AXIS}],\\
&[\texttt{LOOK\_SEGMENT},\texttt{LOOK\_BAR0},\texttt{LOOK\_AXIS},\texttt{LOOK\_SEGMENT}],\\
&[\texttt{LOOK\_SEGMENT},\texttt{LOOK\_BAR1},\texttt{LOOK\_AXIS},\texttt{LOOK\_SEGMENT}],
\end{align}
and, for each \(r\in\mathcal R\),
\begin{align}
&[\texttt{LOOK\_BAR0},\texttt{LOOK\_AXIS},\texttt{LOOK\_SEGMENT},\texttt{REPORT}_r],\\
&[\texttt{LOOK\_BAR1},\texttt{LOOK\_AXIS},\texttt{LOOK\_SEGMENT},\texttt{REPORT}_r],\\
&[\texttt{LOOK\_AXIS},\texttt{LOOK\_SEGMENT},\texttt{REPORT}_r,\texttt{REPORT}_r],\\
&[\texttt{LOOK\_SEGMENT},\texttt{REPORT}_r,\texttt{REPORT}_r,\texttt{REPORT}_r],\\
&[\texttt{REPORT}_r,\texttt{REPORT}_r,\texttt{REPORT}_r,\texttt{REPORT}_r].
\end{align}

For the Fast agent, the exploratory templates operate on the pair-level
gist--anchor--refine sequence:
\begin{align}
&[\texttt{LOOK\_PAIR},\texttt{LOOK\_AXIS},\texttt{LOOK\_SEGMENT},\texttt{LOOK\_AXIS}],\\
&[\texttt{LOOK\_PAIR},\texttt{LOOK\_AXIS},\texttt{LOOK\_SEGMENT},\texttt{LOOK\_PAIR}],\\
&[\texttt{LOOK\_AXIS},\texttt{LOOK\_SEGMENT},\texttt{LOOK\_PAIR},\texttt{LOOK\_AXIS}],\\
&[\texttt{LOOK\_SEGMENT},\texttt{LOOK\_AXIS},\texttt{LOOK\_SEGMENT},\texttt{LOOK\_PAIR}],
\end{align}
and, for each \(r\in\mathcal R\),
\begin{align}
&[\texttt{LOOK\_PAIR},\texttt{LOOK\_AXIS},\texttt{LOOK\_SEGMENT},\texttt{REPORT}_r],\\
&[\texttt{LOOK\_PAIR},\texttt{LOOK\_AXIS},\texttt{REPORT}_r,\texttt{REPORT}_r],\\
&[\texttt{LOOK\_AXIS},\texttt{LOOK\_SEGMENT},\texttt{REPORT}_r,\texttt{REPORT}_r],\\
&[\texttt{LOOK\_SEGMENT},\texttt{REPORT}_r,\texttt{REPORT}_r,\texttt{REPORT}_r],\\
&[\texttt{REPORT}_r,\texttt{REPORT}_r,\texttt{REPORT}_r,\texttt{REPORT}_r].
\end{align}

\end{document}